\documentclass[twocolumn]{aastex631}
\usepackage{longtable}

\begin{document}

\title{Utilizing Machine Learning to Predict Host Stars and the Key Elemental Abundances of Small Planets}

\author[0000-0002-8483-5632]{Amílcar R. Torres-Quijano}
\affiliation{University of Texas at San Antonio, Physics and Astronomy Dept., One UTSA Circle, San Antonio, TX, 78249, USA}

\author[0000-0003-0595-5132]{Natalie R.\ Hinkel}
\affiliation{Louisiana State University, Department of Physics and Astronomy, 202 Nicholson Hall, Baton Rouge, LA 70803, USA}

%\collaboration{20}{(AAS Journals Data Editors)}

\author[0000-0001-5563-6987]{Caleb H. Wheeler III}
\affiliation{LIGO Livingston Observatory, Livingston, LA 70754, USA}
%\affiliation{AAS Journals Associate Editor-in-Chief}

\author[0000-0003-1705-5991]{Patrick A. Young}
\affiliation{School of Earth and Space Exploration, Arizona State University, Tempe, AZ 85287, USA}

\author[0000-0002-9089-0136]{Luan Ghezzi}
\affiliation{Observat\'orio do Valongo, Universidade Federal do Rio de Janeiro, Ladeira do Pedro Ant\^onio, 43, Rio de Janeiro, RJ 20080-090, Brazil}

\author{Augusto P. Baldo}
\affiliation{Observat\'orio do Valongo, Universidade Federal do Rio de Janeiro, Ladeira do Pedro Ant\^onio, 43, Rio de Janeiro, RJ 20080-090, Brazil}

\begin{abstract}

Stars and their associated planets originate from the same cloud of gas and dust, making a star's elemental composition a valuable indicator for indirectly studying planetary compositions. While the connection between a star's iron (Fe) abundance and the presence of giant exoplanets is established \citep[e.g.][]{Gonzalez1997,Fischer05}, the relationship with small planets remains unclear. The elements Mg, Si, and Fe are important in forming small planets. Employing machine learning algorithms like XGBoost, trained on the abundances \citep[e.g., the Hypatia Catalog, ][]{Hinkel2014} of known exoplanet-hosting stars (NASA Exoplanet Archive), allows us to determine significant ``features" (abundances or molar ratios) that may indicate the presence of small planets. We test on three groups of exoplanets: (a) all small, R$_{P}$ $<$ 3.5 $R_{\oplus}$, (b) sub-Neptunes, 2.0 $R_{\oplus}$ $<$ R$_{P}$ $<$ 3.5 $R_{\oplus}$, and (c) super-Earths, 1.0 $R_{\oplus}$ $<$ R$_{P}$ $<$ 2.0 $R_{\oplus}$ -- each subdivided into 7 ensembles to test different combinations of features. We created a list of stars with $\geq90\%$ probability of hosting small planets across all ensembles and experiments (``overlap stars"). We found abundance trends for stars hosting small planets, possibly indicating star-planet chemical interplay during formation. We also found that Na and V are key features regardless of planetary radii. We expect our results to underscore the importance of elements in exoplanet formation and machine learning's role in target selection for future NASA missions: e.g., the James Webb Space Telescope (JWST), Nancy Grace Roman Space Telescope (NGRST), Habitable Worlds Observatory (HWO) -- all of which are aimed at small planet detection.

\end{abstract}


\section{Introduction} \label{sec:intro}
Stars and their planets formed from the same molecular cloud core. The rotating disk of gas and dust that surrounds a young star can accrete into larger objects which can lead to the formation of planets \citep[e.g.][]{Liu2020}. For example, Mg, Si, and Fe can condense into rocky material, which is important in forming small planets. Since a star and its orbiting planets form from the same giant molecular cloud at roughly the same time, we can use stellar abundances as a proxy for studying small planet composition. This is a necessarily indirect method since it is incredibly difficult to directly measure an exoplanet’s interior composition \citep[the exception being polluted white dwarfs, e.g.,][]{Bonsor2011,Xu2019}.  

While there are a variety of definitions of a small planet, we build upon work done by \cite{Bergsten2022} who define small planets as those with radius R$_P$ $<$ 3.5 $R_{\oplus}$. The cutoff at $R_{P}$ $<$ 3.5 $R_{\oplus}$ is due to a noticeable drop in planet occurrence for the \cite{Bergsten2022} dataset, which was composed entirely of Kepler small planet detections. Including Kepler demographic data when defining these small planet regimes is critical for understanding the biases of the observed data and their implications on our results, as most ($\approx91.64\%$) of the small planet data within the NASA Exoplanet Archive\footnote{https://exoplanetarchive.ipac.caltech.edu/} have been Kepler/K2 mission discoveries.
\cite{Bergsten2022} further define super-Earths as planets between 1 - 2 $R_{\oplus}$ and sub-Neptunes as planets between 2 - 3.5 $R_{\oplus}$, which we also adopt. These radius cutoffs were chosen to include the demographic characteristics seen when observing the radius valley for the Kepler data \citep{Fulton2017, Martinez2019, Bergsten2022, Loaiza-Tacuri2023}. 

A connection has been established between a star's iron content (or [Fe/H], see an explanation of stellar abundance dex notation in \citealt{Hinkel2022}) and the existence of giant exoplanets \citep[e.g.][]{Gonzalez1997,Fischer05}, but a similar link is not yet evident for small planets. While a star's [Fe/H] is commonly used as a proxy for stellar metallicity \citep[e.g.][]{Fischer05,Bashi2019}, other elements are important for planetary formation \citep{Gonzalez2009, Helled2014}. For example, $\alpha$-elements: Mg, Si, in addition to Fe are critical in the formation of rocky material where Mg, Si, and Fe make up more than $\>90\%$ by mass of chondrites, Earth and the other rocky planets in our Solar System \citep[excluding Mercury,][]{McDonough2003,Unterborn2023}. The combinations of these elements also form minerals, such as olivine $\textrm{(Mg,Fe)}_{2} \textrm{SiO}_{4}$, which is a primary component of the upper mantle of the Earth \citep{Smyth2006}. 

Multiple studies have been conducted \citep[e.g.][]{Schulze2021,Unterborn2023} to understand the composition of a planet from the chemical abundance of the host star. \cite{Schulze2021} began their analysis with the assumption that the mass and radius of an exoplanet is compatible with the model of a planet that consists only of an Fe core and a silicate mantle (without Fe) that matches the proportions of their host stars photospheric Fe/Mg and Si/Mg abundance ratios. The authors test the hypothesis by comparing two calculations of a planet's core mass fraction (CMF) based on a planet's mass and radius. They calculate the CMF, which results in the average density of the planet (CMF$_{\rho}$) and compare it to the CMF as predicted by the Mg, Si, and Fe abundances of the planet's host star (CMF$_{star}$). If the determinations of both CMF$_{\rho}$ and CMF$_{star}$ differed, then their hypothesis would be rejected. \cite{Schulze2021} examined 11 exoplanets and determined that the mass and radius of Kepler 107 c do not allow it to be described as a likely rocky planet with the same relative elemental abundances of rock-building elements as its host star. For HD 219134 c, they found that an Fe abundance 40$\%$ greater than the CMF$_{star}$ indicates that the planet would have different rock-building element composition compared to the host star, such as a ``super-Mercury" class of exoplanets, possibly due to physical processes. The authors were also able to identify low-density planets (e.g., 55 Cnc e) where the CMF$_{\rho}$ was 50$\%$ less than was predicted by the CMF$_{star}$. They conclude that these planets show that small planets below the radius gap can have varied compositions and may not solely rocky planets. However, they could not confirm their initial hypothesis based on mass or radius for more than half of their sample due to the large associated measurement errors. 

Similarly, \cite{Unterborn2023} sampled seven planets with radii between 1.2-1.6 R$_{\oplus}$, with available host star chemical abundances to test suitable proxies for planetary composition. When authors compared the CMF$_{\rho}$ and CMF$_{star}$ for their seven-planet sample, they found that the composition of three planets (Kepler-102 d, Kepler-105 c,
and Kepler-406 b) deviated from their host's composition. Due to the planets' close orbit and higher than expected density, these planets were confirmed as super-Mercuries. The compositions of the remaining four planets in their sample (Kepler-99 b, Kepler-78 b, Kepler-36 b, and Kepler-93 b) were not statistically distinguishable from those of their hosts, either because the planets had significant gaseous envelopes or because of uncertainties in the planet's mass, radius, and stellar composition. They conclude that as uncertainties in mass, radius, and stellar abundances decrease, the authors would be able to constrain further the hypothesis that a star's stellar abundance can be used as a proxy to study planetary composition. 

In both of these studies, there were situations in which the star-planet link was not established. They indicate that this is likely due to physical processes that disrupt their initial compositions, such as impacts, anomalous host star compositions, or processes that are still unknown. Yet, these studies allow us to develop more constrained planetary models and potentially clarify how formation processes of small planets are expressed in their composition.

\cite{Thiabaud2015a} also set out to explore the chemical relationship between a star and its orbiting planets. They utilized a planet formation model \citep{Alibert2013} combined with a chemical model \citep{Marboeuf2014a,Marboeuf2014b,Thiabaud2014,Thiabaud2015b} to determine the elemental composition of the modeled planets. The planets orbited 18 stars of solar mass with varying molar ratios of Fe/Si = 1.34 -- 2.81 and Mg/Si = 0.66 -- 1.85. The planets were categorized as rocky, icy, or giant -- in all three categories, they found that the Mg/Si and Fe/Si composition of the modeled planets mirrored the composition of their host stars in a 1:1 relationship. For the icy and giant planets, they note that this was an expected result since the refractory elements would condense within their formation regions. The Mg/Si and Fe/Si molar ratios could vary for rocky planets due to either the location of where the planet formed or where it settled after migration. These variations could lead to planets with different Mg/Si and Fe/Si compositions compared to their hosts. The authors showed how molar ratios can help establish a link between the composition and formation processes of stars and planets.

In this paper, we use stellar abundances available from the Hypatia Catalog \citep{Hinkel2014} combined with data from the NASA Exoplanet Archive within the framework of a ``supervised classifier" machine learning algorithm to identify which elements may help determine the presence of small planets and produce a target list of potential stellar hosts. In Section \ref{sec:data}, we define the data sets used for training and prediction, the subdivision of the experiments, the features analyzed, and highlight potential data biases. We give an overview of the machine learning algorithm (XGBoost) in Section \ref{sec:algorithm} and define the training sample. Section \ref{sec:results} highlights the important results obtained for the study, such as the relevant elements (or features), and highlights the stars that have a high likelihood of hosting small planets. Finally, we interpret the significance of the features with highest importance, discuss how molar ratios can help us understand the distribution of minerals within a system, and highlight the usefulness of machine learning in the creation of target lists for future science missions in Section \ref{sec:discussion}.

\section{Data} \label{sec:data}
\subsection{Stellar Elemental Abundances}\label{subsec:abundances}
We obtain the stellar elemental abundance data for this study from the Hypatia Catalog\footnote{\url{www.hypatiacatalog.com}} \citep{Hinkel2014}. It contains the abundance data for more than 100 elements and species for $>$11,000 FGKM-type stars within 500 pc of the Sun. Of these stars, $>$1,400 are known exoplanet hosts. We required a sufficiently large sample of stars ($>$200) with recorded stellar abundances as per \cite{Hinkel2019} to create training and prediction datasets. We additionally required that each star have at least $50\%$ of its relevant abundance values be measured (as opposed to missing or ``null" values) to avoid potentially biased results due to the inclusion of missing values, which are further discussed in Section \ref{subsec:nullvalues}. After removing those stars with $<$50\% recorded abundances (which included all M-stars), we obtained a dataset with 10,178 stars.

Stellar abundances are usually denoted in ``decadic logarithmic units" (dex). This system is a base 10 logarithmic scale, which helps compare abundance values which often have large dynamic ranges. Dex notation defines the abundance of hydrogen in a star as 12, setting the maximum of the scale, with all other abundances having absolute values less than 12. The dex notation also commonly normalizes stellar abundances with respect to solar abundances for given elements, indicated with square brackets. For example, element (Q) is often examined with respect to H or Fe: [Q/H] or [Q/Fe], respectively \citep{Hinkel2022}. While dex notation is the primary system for expressing chemical abundance data in the astrophysics community, converting stellar abundances to molar ratios, expressed as Q/Fe, without solar normalization provides the same data in a format that ties directly to the proportions and chemical reactants needed to for determination of mineralogical compositions of planetary interiors \citep{Hinkel2018}.

We focus on the abundances for 16 elements: C, O, Na, Mg, Al, Si, Ca, Sc, Ti, V, Cr, Mn, Fe, Co, Ni, and Y. The elements chosen for our analysis fall within the lithophile or siderophile groups. Lithophiles (Na, Mg, Al, Si, Ca, Sc, Ti, V, Mn, Y) are elements that bond with oxygen and form oxidized minerals, while siderophiles (Cr, Co, Ni) commonly alloy with Fe. The volatile element O was chosen due to its bonding nature with elements such as Mg, Si, and Fe since they form minerals such as olivine and pyroxenes, which are important in creating planetary interiors. The volatile elements (C and O) were also included based on the potential for a planet to possess volatile envelopes. For example, both of these elements can be present in a small planet atmosphere in oxidized (e.g. CO$_{2}$) or reduced forms \citep[e.g., CO,][]{Gaillard2022}. As for molar ratios, these are also studied due to their utility in determining the properties of planetary interiors. As such, we chose to study molar ratios with respect to Mg, Si and O, in our analysis. We go into more detail about how these elements are combined into ensembles in Section \ref{subsec:ensembles}. 

\subsection{Exoplanet Host Stars}\label{s.exoplanethosts}

We obtain small planet data from the NASA Exoplanet Archive, an online catalog that collects publicly available data regarding the search and characterization of exoplanets and currently contains information for $>$5,500 confirmed exoplanets. For our analysis, we remove any planets that have radii R$_{P}>$ 3.5 R$_{\oplus}$. Furthermore, if the host star has multiple known planetary companions, we choose to keep the largest planet in the system (below the radius cutoff) since the largest planets are most easily observable and have the stronger chemical interplay with the host star. This decision was made to remove duplicating a planet hosting star within the analysis, as the multiple instances of the stellar abundance values would skew the results. Finally, we cross match these small planet hosts to the stars available in the Hypatia Catalog to ensure that they have stellar abundances. This gives us a total of 479 stars with known planets that fall within 0.0 $<$ R$_{P}<$ 3.5 R$_{\oplus}$. Furthermore, this results in 211 super-Earth planets (1 - 2 $R_{\oplus}$) and 219 sub-Neptune planets (2 - 3.5 $R_{\oplus}$) -- representing the respective training samples (see Section \ref{sec:algorithm}) for each experiment (described below). Removing these known planet hosts from our overall dataset of 10,178 stars, we end up with 9,698 stars without known planets as our prediction sample (see Section \ref{sec:algorithm}). By combining this dataset, alongside the stellar abundance data from the Hypatia Catalog, we can obtain a full overview of a system: the star, its abundances, and any planets that may orbit it.

Separating small planets into categories helps to determine whether certain elements are more important in determining the presence of a specific exoplanet class. For example, sub-Neptunes might be more heavily influenced by volatiles (C, O), while super-Earths might be more heavily influenced by elements important in forming planetary interiors (Mg, Si, Fe).

We separate our study into three experiments based on planetary radii, as per \cite{Bergsten2022}, where a minimum of 200 planets are necessary to successfully train the algorithm (see Section \ref{sec:algorithm}). 

\begin{description}
    \item[Experiment 1] Small Planets (479 planets with R$_{P} <$ 3.5 R$_{\oplus}$); Stellar host distribution: 79 F-types, 294 G-types, 106 K-types
    \item[Experiment 2] Sub-Neptunes (219 planets with 2.0 R$_{\oplus} <$ R$_{P} <$ 3.5 R$_{\oplus}$); Stellar host distribution: 28 F-types, 131 G-types, 60 K-types
    \item[Experiment 3] Super-Earths (211 planets with 1.0 R$_{\oplus} <$ R$_{P} <$ 2.0 R$_{\oplus}$); Stellar host distribution: 42 F-types, 135 G-types, 34 K-types
\end{description}
\noindent
Experiment 1 allows us to analyze which general features are the most important in determining the presence of a small planet. Experiment 2 tests to see if certain elements may have a greater impact on determining sub-Neptunes' presence, possibly due to the volatile envelopes these planets might harbor. Additionally, 
Experiment 3 tests if the elements important to the formation of planetary interiors are the most important in detecting the presence of a super-Earth.

\begin{table*}[t]
\centering
\begin{tabular}{|c|c|c|c|c|} 
 \hline
 Ensemble & Features & Experiment 1 & Experiment 2 & Experiment 3 \\ [0.5ex] 
 \hline\hline
 1 & C, O, Mg, Si, Ti & 4,767 & 4,767 & 9,698 \\
 2 & C, O, Mg, Si, Ti, Fe & 4,767 & 4,767 & 9,698 \\
 3 & Volatiles + Lithophiles + Siderophile & 2,486 & 2,486 & 9,698 \\
 4 & Volatiles + Lithophiles + Siderophile + Fe & 2,486 & 2,486 & 9,698 \\
 5 & Si/Mg, Ti/Mg, Fe/Mg, C/Mg, Ca/Mg, O/Mg & 4,765 & 4,765 & 9,698 \\
 6 & Mg/Si, Ti/Si, Fe/Si, Ca/Si, C/Si, O/Si & 4,765 & 4,765 & 9,698 \\
 7 & Si/O, Ti/O, Fe/O, Ca/O, Mg/O, C/O & 4,765 & 4,765 & 9,698 \\
 \hline
\end{tabular}
\caption{List of each tested ensemble by number (column 1), features tested (column 2), and the number of stars available for prediction for each experiment (columns 3-5, respectively). We note that volalites here include C, O; lithophiles include Na, Mg, Al, Si, Ca, Sc, Ti, V, Mn, Y; and siderophiles are Cr, Co, Ni. See Section \ref{sec:data} for more details.}
\label{tab:ens-list}
\end{table*}

For the stellar hosts, the training sample for each experiment have effective temperatures ranging from $\sim$3900-6900 K, indicating a range of F-, G-, and K-type stars. Across all experiments, G-type stars consistently outnumber F- and K-type stars in the training samples. The F-type stars are the least common in the training samples of Experiments 1 and 2, whereas K-type stars are the least common in Experiment 3 by a small margin.

All 479 of the planets within our training sample were discovered via the transit method. This planet detection method, which is common for planets with R$_{P}<$3.5$_{\bigoplus}$, has inherent biases which can affect our data. The transit method is more likely to detect planets with a smaller orbital semi-major axis as well as a larger planetary radii that can obscure more light when passing in front of their host star.  An additional bias for transit detections is that planets with shorter orbital periods are easier to confirm smaller observation campaigns. As such, the results for our analysis could heavily reflect the important features necessary for detecting small planets of larger radii that are at short orbital periods ($<$ 100 days) from their host star. While a transit analysis is sufficient to obtain a planet's radius, determining mass requires additional methodologies (e.g., radial velocity).

\subsection{Element Ensembles}\label{subsec:ensembles}
We further separate each of our three experiments into seven ensembles, shown in Table \ref{tab:ens-list}. These ensembles are different combinations of either element abundance values expressed in dex notation or molar ratios. The lack, or inclusion, of Fe is used to test the trend of increased small-planet occurrence rates around Fe-poor $\alpha$-rich stars \citep[e.g.][]{Brewer2018, Bashi2019}. Ensembles 1 and 2 tests the influence of individual stellar $\alpha$-element abundances and Fe to predict the presence of small planets. These ensembles also include C and O to test the influence of volatile elements, which could be important in determining the presence of sub-Neptunes or planets composed primarily of water and volatiles (water worlds). Ensembles 3 and 4 include the lithophiles and siderophiles to test the significance of elements known to form minerals in determining the presence of a planet.

The molar ratio ensembles (5-7) were selected to obtain further insight into the specific proportions of elements necessary to determine the presence of a planet. Differences in the values for a molar ratio can indicate the availability of an element to form compounds and minerals. For example, an Mg/Si ratio $<$ 1.0 indicates that Mg will be present primarily in pyroxene while feldspars will obtain most of the additional Si \citep{Bond2010}. Therefore, Mg/Si ratios can help us understand the distribution of silicates for a planetary system \citep{Bond2010, Thiabaud2015a}. Meanwhile, Fe/Mg ratios can influence a planetary core's size, affecting heat flow from core to mantle \citep{Hinkel2018}. We also look at molar ratios with respect to O due to the bonding nature of O with elements such as Mg and Si for additional insights into the distribution of minerals.

In order for XGBoost to make predictions on a given star, the star must have values recorded for all of the elements in the abundances or molar ratios within the ensemble. The number of available stars predicted upon for each ensemble within each experiment is given in Table \ref{tab:ens-list}. We note that Experiment 3 is an exception, where null abundances were included (see Section \ref{subsec:nullvalues}).

\section{Algorithm} \label{sec:algorithm}

Machine learning (ML) can be separated into supervised and unsupervised learning. Supervised training pairs input data (features) with known output data to develop a model. Supervised classifiers (logistic regression) are a further distinction, where inputs are mapped to a category or label output values. For illustration, consider a hypothetical movie streaming service that wants to provide movie suggestions to a user based on previously watched material. Once a user watches various movies or shows (the features) and provides feedback in binary ``likes," that pairs movie keywords to a user, thus producing the training set. A sufficiently large training set allows the algorithm to determine the relationships between the watched material (i.e., what genres) and what unwatched material exists to recommend. In summary, supervised classifier ML algorithms use the labeled training data (the training sample) to derive general relationships and apply feature-based predictions to unseen or future data (the prediction sample).

Supervised classifier algorithms can be further subdivided, and for the present work, we use a gradient boosting supervised classifier XGBoost \citep{ChenandGuestrin2016}. Gradient boosting constructs a prediction model by combining weaker models, such as decision trees, that make minimal assumptions about the input data. Thus, a prediction model can be constructed by combining multiple decision trees trained simultaneously (parallel tree boosting). These boosted decision trees are trained to correct mistakes present in an earlier tree of the series. XGBoost is an implementation of the gradient boost ML algorithm which provides parallel tree boosting to solve data science problems \citep{ChenandGuestrin2016}.

In our study, XGBoost follows a similar approach to the streaming service analogy. We use the stellar abundances of known exoplanet hosts (479, 219, and 211 for the three experiments, respectively) as the training sample. Our prediction sample contains 9,698 stars currently not known to host small planets. In order to treat the imbalance in sample sizes, we chose to undersample the prediction sample \citep{Feng21}. Therefore, when executing the algorithm for each experiment, we chose a random sub-sample of 200 stars from each set due to significant differences in population size between the prediction and the training sample. The XGBoost algorithm tests for patterns of stellar abundances of known planet hosts in the training sample to develop a model. The model provides an output prediction of the presence of small planets based on the input of stellar abundances. With the model, we can then apply the algorithm to the prediction sample or the abundances of those stars not known to host exoplanets. This is the same technique implemented by \cite{Hinkel2019} to determine the important elements that predict the presence of a giant planet host. Each ensemble is run 25 times, meaning there are 25 initializations of XGBoost per ensemble. Every XGBoost initialization is repeated for 3,000 iterations\footnote{An initialization refers to the start of a computational loop for XGBoost, while an iteration refers to a nested computational loop that repeats that instance of XGBoost.}, and each iteration chooses a new subgroup of 200 stars \citep{Hinkel2019}. This allows us to produce both (a) a probability for a non-host star to host a small planet, calculated from over a thousand iterations, and (b) determine the importance of a feature in predicting the presence of an exoplanet. The probability for a star to host a small planet is calculated by dividing the total number of instances that a star is confidently predicted (Pred) as a planet host by XGBoost by the total number of instances each star was tested (Sampled) by XGBoost: Prob = Pred/Sampled \citep{Hinkel2019}. The overarching code, which employs XGBoost and includes a \texttt{params.yaml} file with all of the employed classifier parameters, is known as PlanetPrediction and is publicly available at \href{https://www.github.com/HypatiaOrg/planetPrediction/tree/rocky}{github.com/HypatiaOrg/planetPrediction/tree/rocky}.

\begin{figure*}[t]
\gridline{\fig{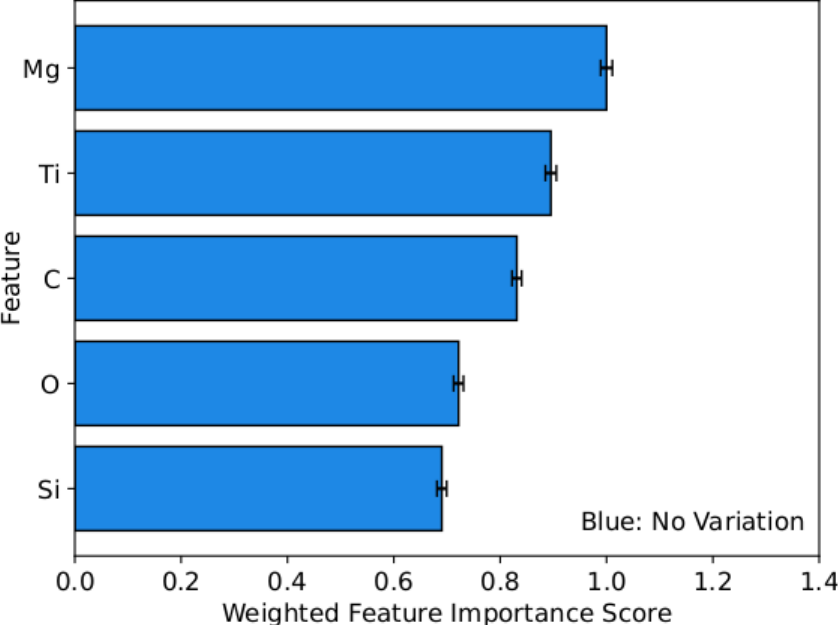}{0.5\textwidth}{(a) Ensemble 1}
          \fig{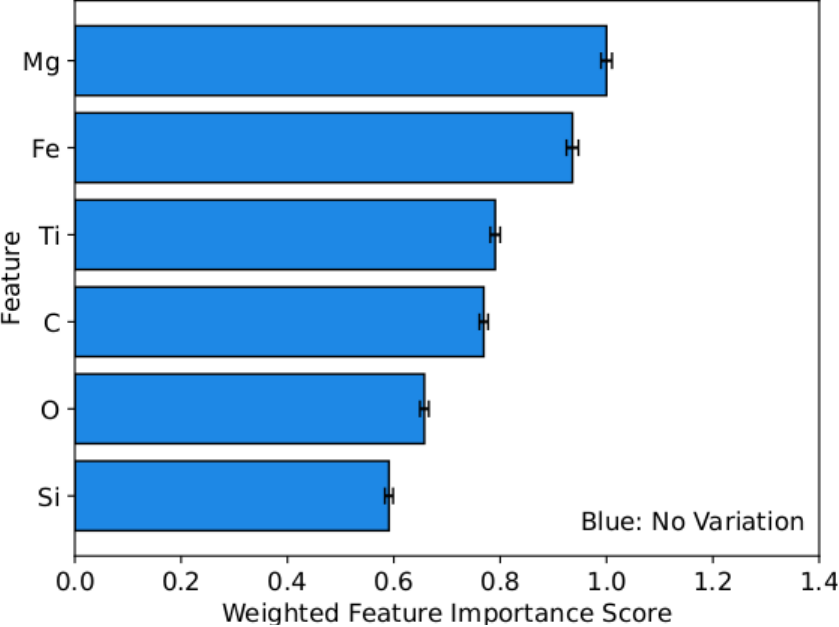}{0.5\textwidth}{(b) Ensemble 2}
          }
\gridline{\fig{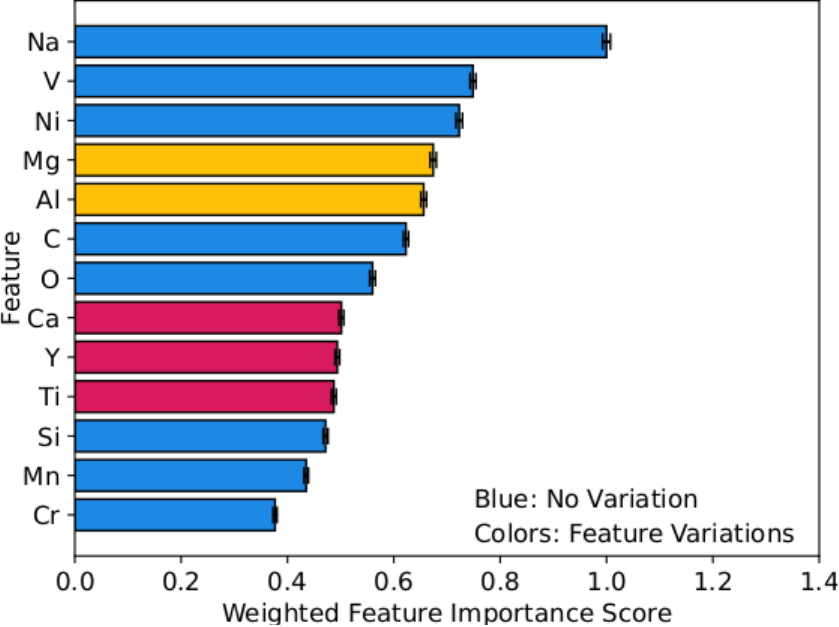}{0.5\textwidth}{(c) Ensemble 3}
          \fig{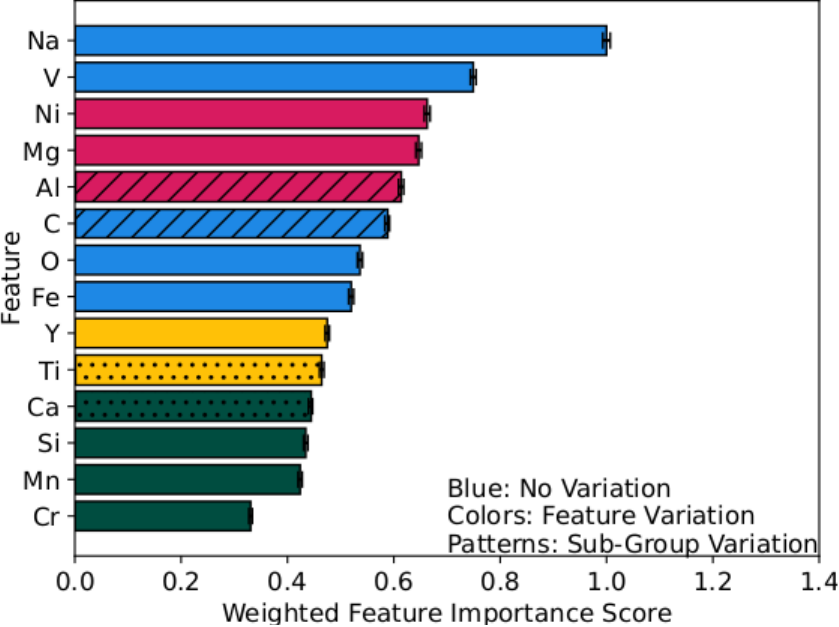}{0.5\textwidth}{(d) Ensemble 4}
          }
\caption{Feature importance plots obtained for the ensembles in Experiment 1  (Small Planets). Because the PlanetPrediction algorithm performs thousands of iterations over which the feature importance scores may vary. We have employed a color coding and pattern schema to identify feature importance score variations. Blue coloration denotes that a particular element did not have any variations in importance score across initializations during an ensemble run. In contrast, colors other than blue indicate that those elements varied in feature importance. The hatching patterns within the lower panels (c) and (d) indicate additional smaller (sub-group) variations that occurred within the ensembles.
\label{fig:feature-importances-ex1}}
\end{figure*}

\subsection{The Golden Set} \label{subsec:golden}
We established a ``golden set" of known exoplanet host stars to assess the algorithm's accuracy. The golden set stars were not used for training the algorithm but rather were reserved for testing its predictive capabilities. During each iteration, the algorithm selects 10 stars known to host small planets at random (golden set stars) -- this number was chosen to ensure the sample remained $>$200 planet hosting stars, see Section \ref{s.exoplanethosts}. It then includes them in the prediction sample without prior knowledge that the golden set stars are known small planet hosts. The prediction model is then applied to golden set stars as a way to test ``true-positives", resulting in a list of probabilities for hosting a small planet. This allows us to test the fidelity of the prediction model at the end of each iteration.

As each initialization of XGBoost performed thousands of iterations, we can derive statistical metrics for the golden set by aggregating the probabilities calculated for each group of golden set stars. For Experiment 1, we had a total of 1,475 stars chosen for the golden set with an average predicted value of 80.25$\%$. Experiment 2 had 1,483 total stars used within the golden set, and their average predicted value was 79.40$\%$. Lastly, for Experiment 3, the average predicted value was 90.26$\%$ with 1,482 total stars used in the golden set (see Section \ref{subsec:3}). In total, we obtain an 83.30$\%$ overall prediction score for the small planet hosts within the samples of the golden set. This high percentage rate shows that the algorithm is effectively predicting small planet host stars. 

\cite{Tamayo2016} note that there exists a probability threshold wherein there is a balance between how useful (precision) and how complete the results are (recall or sensitivity). For this reason, we choose to use a conservative baseline wherein we define any star with a $\geq90\%$ prediction score as a likely planet host due to our models' accuracy in predicting small planet hosts. A star with a prediction probability exceeding this baseline are more likely to have an undiscovered small planet that can be confirmed through future observations.

\begin{figure*}[t]
\gridline{\fig{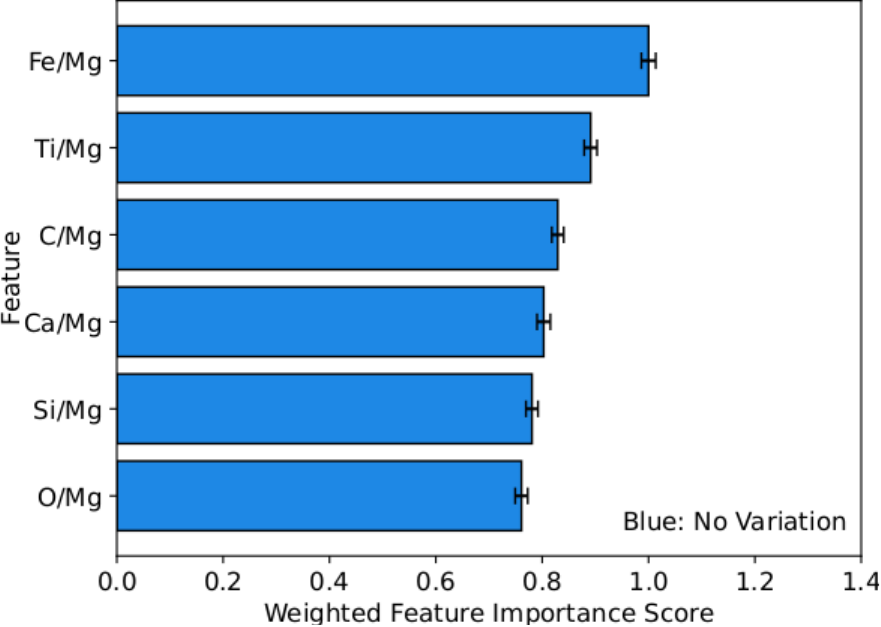}{0.5\textwidth}{(a) Ensemble 5}
          \fig{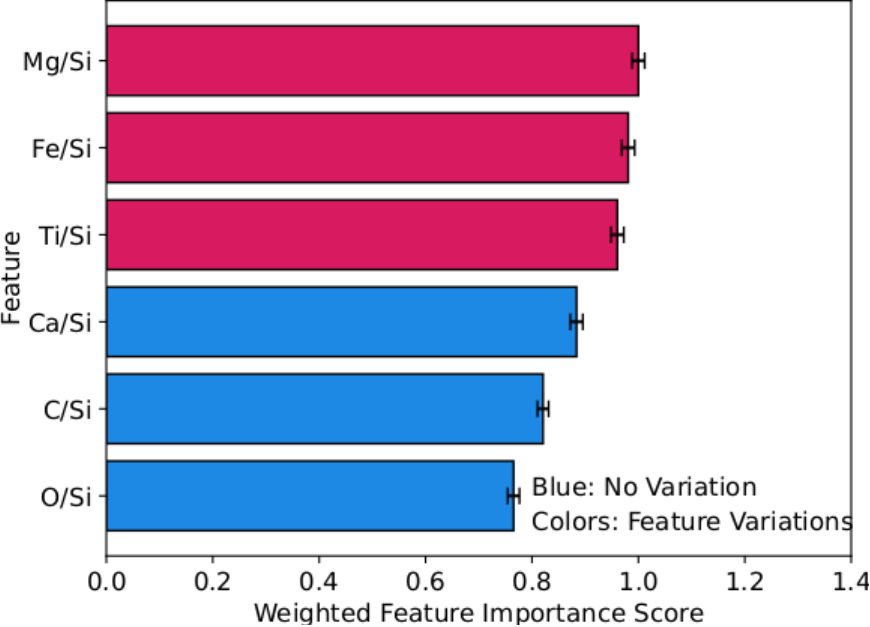}{0.5\textwidth}{(b) Ensemble 6}
          }
\gridline{\fig{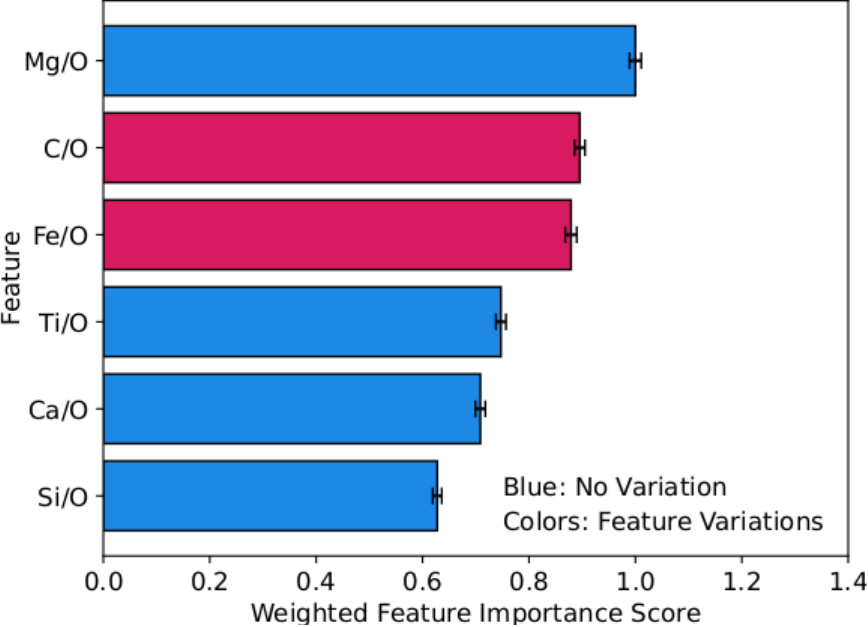}{0.5\textwidth}{(c) Ensemble 7}
          }
\caption{Similar to Fig. \ref{fig:feature-importances-ex1} but for the molar ratios tested in Experiment 1. Given that Fe/Mg is the most important features for Ensemble 5, this implies that the size of a small planet's core (Fe/Mg) could be an important factor in determining the presence of a small planet (see Section \ref{sec:data}). The molar ratios of Mg/Si and Fe/Si are often the most important features in Ensemble 6, while Mg/O is the most significant in determining the presence of a small planet in Ensemble 7.
\label{fig:molar-ratios-ex1}}
\end{figure*}

\subsection{Validation of Ensemble Sizes}

The ensembles for a given experiment can be set up with any number of features within them. \cite{Hinkel2019} used $\geq$ 13 elements per ensemble. Since the ensembles within our study use a varying number of elements (5 - 14 elements), we wanted to ensure that the results would be robust for ensembles of smaller size. For this reason, we created validation ensembles with a smaller number (from 3 - 6) of elements to validate the XGBoost prediction model. The stars with high probabilities of hosting a small planet ($\geq$90$\%$) within these validation ensembles were compared with those stars from Experiment 1, which also showed a high ($\geq$90$\%$) probability of hosting a small planet. This allowed us to determine how many stars were shared (shared stars) between both groups. 
When comparing the validation ensemble with three elements for the Experiment 1 stars, we noted that this validation ensemble resulted in the least number of shared stars (26). The same was true when we had ensembles of 4 elements. However, the validation ensemble with 5 or 6 elements resulted in $>$60 shared stars once compared with the Experiment 1 stars. Thus, more elements allow us to obtain a larger number of potential host stars because the algorithm requires more elements to reach the require baseline threshold for small planet host predictions. We used ensembles with $\geq$5 elements to ensure robustness within the XGBoost prediction models.

\subsection{Null Values} \label{subsec:nullvalues}
We consider null values to be stellar abundance data that is unavailable for a given element and and star. If null values are present within the prediction sample, XGBoost can include the null values in the analysis. However, the null values can bias the results because XGBoost uses them to guide the decision trees in a direction that positively predicts a small planet host; XGBoost will favorably predict an exoplanet based on null values. For this reason, stars with many null values are limited in our ensembles to reduce the false high planet prediction scores (see Section \ref{subsec:abundances}). The exception to the exclusion of null values is for Experiment 3, which required additional stars in order to have the requisite number for the training and prediction samples. However, since null values may bias the results, further analysis is necessary when interpreting the results obtained for Experiment 3 (see Section \ref{subsec:3}).

\begin{figure*}[t]
\gridline{\fig{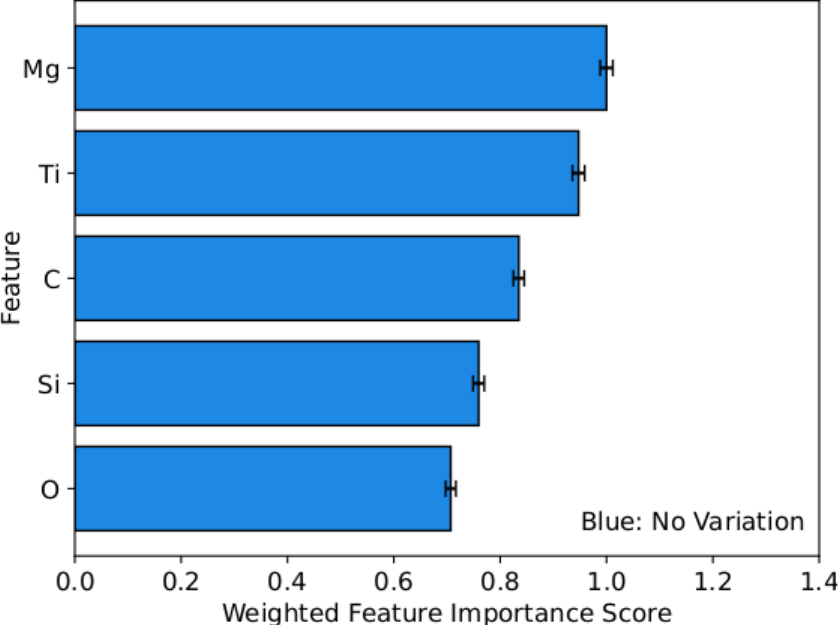}{0.5\textwidth}{(a) Ensemble 1}
          \fig{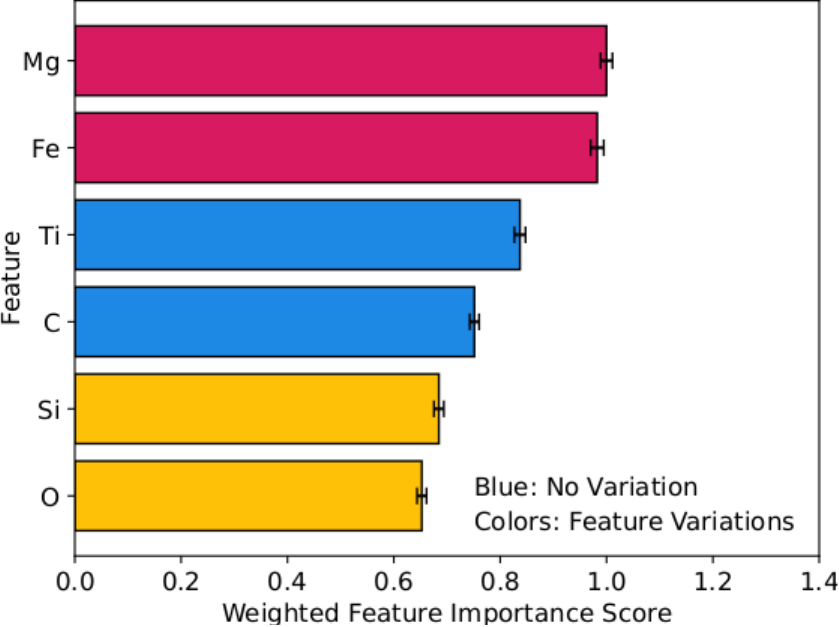}{0.5\textwidth}{(b) Ensemble 2}
          }
\gridline{\fig{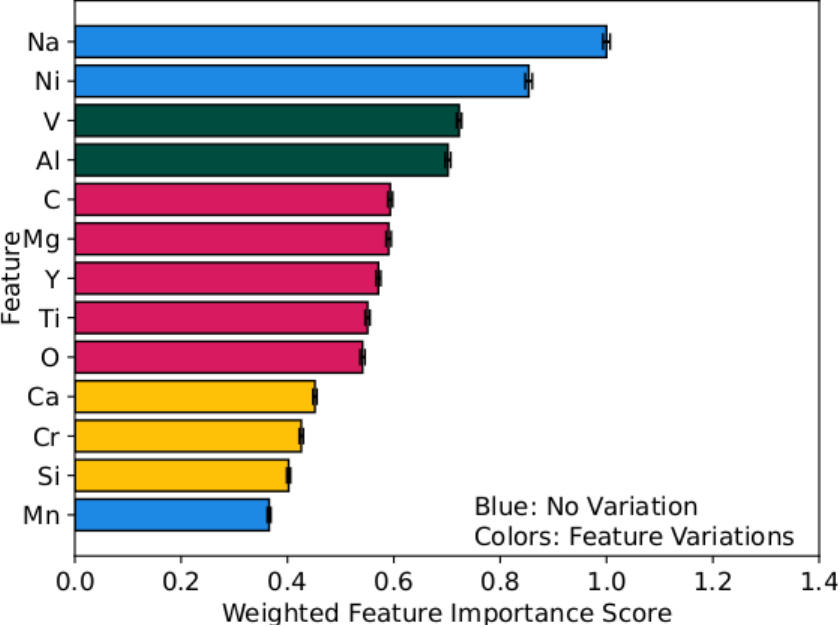}{0.5\textwidth}{(c) Ensemble 3}
          \fig{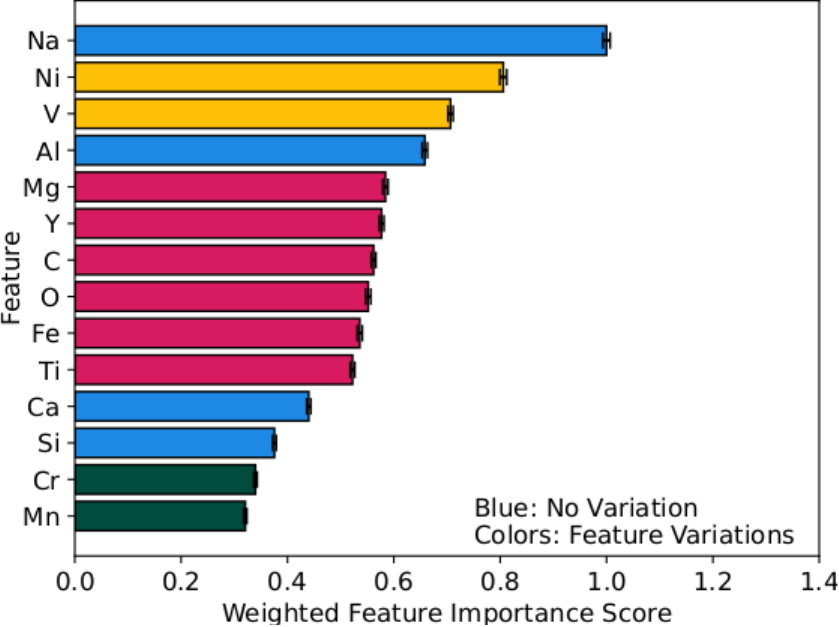}{0.5\textwidth}{(d) Ensemble 4}
          }
\caption{Same as Fig. \ref{fig:feature-importances-ex1} but for the abundance feature importance plots obtained for the Experiment 2 (sub-Neptune) ensembles. Ensembles 1 (a) shows Mg as the most important feature for the sub-Neptunes while Mg and Fe's alternate as the most important feature in Ensemble 2 (b). For Ensemble 3 (c) and Ensemble 4 (d), Na and V -- and to a lesser extent Ni -- are the most important elements for this particular class of planets, similar to Experiment 1.
\label{fig:feature-importances-ex2-part1}}
\end{figure*}

\section{Results} \label{sec:results}

\subsection{Feature Importances} \label{subsec:features}
When starting with a sizable dataset, the data's organization may appear random, leading to the dataset being viewed as having high entropy. This entropy is lowered when the data is split into smaller subsets via similar properties (features), resulting in groups of analogous stars. While a variety of physical and chemical properties could be as different features, our study uses elemental abundances or molar ratios as features that separate the prediction sample into stars into two categories: those less likely to host a small planet and those more likely small planet hosts. A feature that is used more than another during decision tree selection, thus lowering entropy, subsequently increases that feature's importance score \citep{Hinkel2019}. 

The feature importance score is calculated independently for each feature. As discussed in Section \ref{sec:algorithm}, there are 3,000 iterations per XGBoost initialization, with 1,000 decision trees for each iteration. The feature importance score is determined by how much each feature's split point enhances the overall performance measure. Specifically, our model's performance is measured by the likelihood that a star could be a small planet host based on the training set. The importance score for each feature is weighted by the number of observations in each leaf node of a tree and then averaged across all decision trees in the prediction model \citep{Hinkel2019, ChenandGuestrin2016}.

For each ensemble, which initializes XGBoost 25 unique times and iterates 3000 times for each initialization, the PlanetPrediction algorithm produces 25 feature importance plots. Each ensemble has a huge amount of output data, including feature importance scores that may vary between initializations. To illustrate these variations, we employ a color and pattern scheme that allows us to visualize any of the feature importance score changes (see Fig. \ref{fig:feature-importances-ex1} for Experiment 1). Blue indicates no variations were observed for these features, while all other colors show variations between features. Overlaid patterns (hatching) indicate additional (yet smaller) variations occurred for those specific features. 

\begin{figure*}[t]
\gridline{\fig{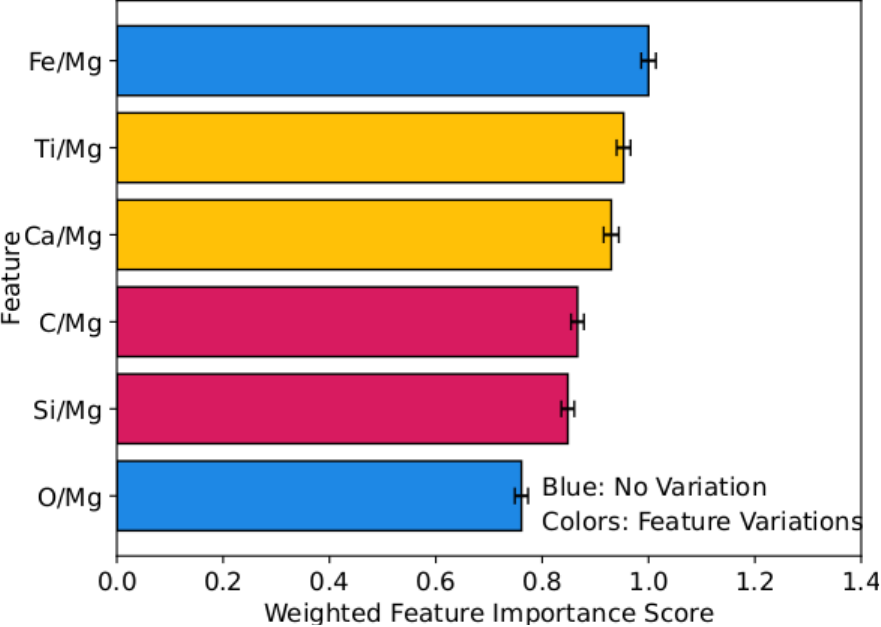}{0.5\textwidth}{(a) Ensemble 5}
          \fig{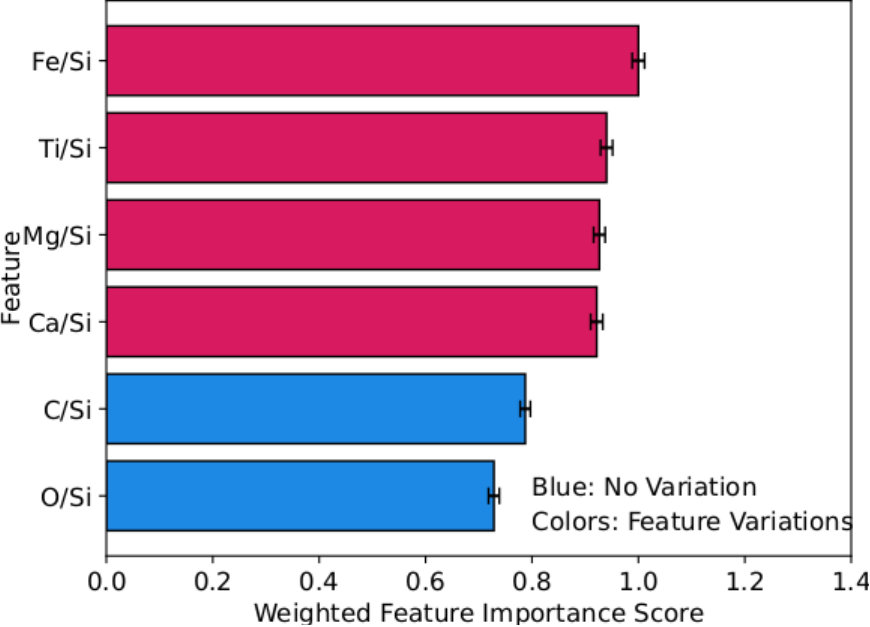}{0.5\textwidth}{(b) Ensemble 6}
          }
\gridline{\fig{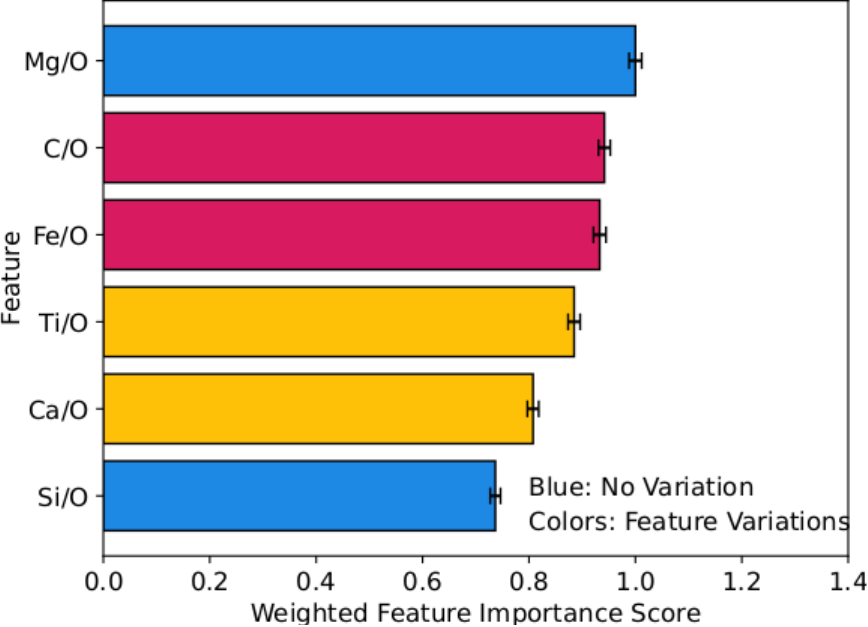}{0.5\textwidth}{(c) Ensemble 7}
          }
\caption{Similar to Fig. \ref{fig:feature-importances-ex1} but for the molar ratio feature importance plots obtained for the Experiment 2 (sub-Neptune) ensembles. Ensemble 5 (a) shows variations in the molar ratios of Ti/Mg and Ca/Mg, as well as C/Mg and Si/Mg. Ensemble 6 (b) exhibits variations among multiple molar ratios (Fe/Si, Ti/Si, Mg/Si, and Ca/Si) to determine the most significant molar ratio in assessing the presence of a sub-Neptune.
\label{fig:feature-importances-ex2-part2}}
\end{figure*}

The error bars for the feature importance scores were calculated via the standard error of the mean for a specific iteration of a boosted decision tree. The errors are with respect to the abundances per iteration -- rather than the variations between the elements (shown in colors). By design, each iteration of the boosted decision tree is segregated from all other decision trees to refrain from contamination of the underlying results (e.g., per each unique subsample of 200 stars from the training sets). Therefore, the error bars do not relate to the feature variations. In general, though, we found that the variations between importance features were generally small, e.g. features often swapped places only with the elements directly above or below them (see Figs. \ref{fig:feature-importances-ex1}-\ref{fig:feature-importances-ex3-part2}). And while variations may have occurred in multiple iterations, creating a larger block of color in the figures, the changes remained modest.

\subsubsection{Experiment 1: Small Planets}
Small planets are defined as those with R$_{P}$ $<$ 3.5 $R_{\oplus}$ \citep{Bergsten2022}. The training sample includes 479 stars, while the number of stars in the prediction sample for each experiment is provided in Table \ref{tab:ens-list}.

We first test the $\alpha$-elements (Ensemble 1) -- which include the volatiles (C, O) and a few lithophiles (Mg, Si, Ti), as shown in Fig. \ref{fig:feature-importances-ex1}a. The element Mg is the most important feature in Ensemble 1. After the addition of Fe, Mg remains the most important element in Ensemble 2. The inclusion of the additional lithophiles and siderophiles (Ensemble 3 and 4, Fig \ref{fig:feature-importances-ex1}c and d, respectively) has Na become the most important feature, and V becomes the element with the second most important feature score. Ensemble 5 (Fig. \ref{fig:molar-ratios-ex1}a) has Fe/Mg as the molar ratio with the highest feature importance score. Ensemble 6 shows that the Mg/Si, Fe/Si, and Ti/Si feature importance scores varied between themselves as the most important feature. Meanwhile, Mg/O is the most prominent feature within Ensemble 7 (Fig. \ref{fig:molar-ratios-ex1}c). These results are further discussed in Section \ref{sec:discussion}.

\subsubsection{Experiment 2: Sub-Neptunes}
Sub-Neptunes include all planets with radii between 2.0 $R_{\oplus}$ $<$ R$_{P}$ $<$ 3.5 $R_{\oplus}$, where there are 219 stars in the training sample. Ensemble 1 (see Fig. \ref{fig:feature-importances-ex2-part1}a) shows that Mg is the most important element in determining the presence of a sub-Neptune while Mg and Fe vary in importance for Ensemble 2 (Fig. \ref{fig:feature-importances-ex2-part1}b). The inclusion of the additional lithophiles and siderophiles (Ensembles 3 and 4, Fig \ref{fig:feature-importances-ex2-part1}c and d, respectively) shows Na as the most important element. The importance score of Ni shifted to a higher significance where it is the second most important element in Ensembles 3 and 4, although V also shows high importance as it varied with Ni in Ensemble 4. In contrast to Experiment 1, Ensemble 6 (Fig. \ref{fig:feature-importances-ex2-part2}b) has multiple molar ratios (Fe/Si, Ti/Si, Mg/Si, and Ca/Si) vary as the most important molar ratio in determining the presence of a sub-Neptune.

\begin{figure*}[t]
\gridline{\fig{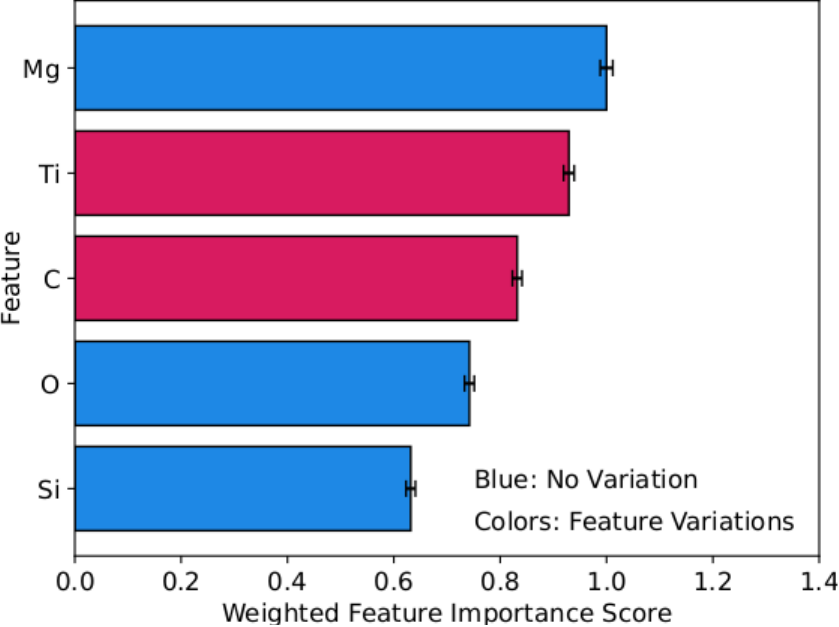}{0.5\textwidth}{(a) Ensemble 1}
          \fig{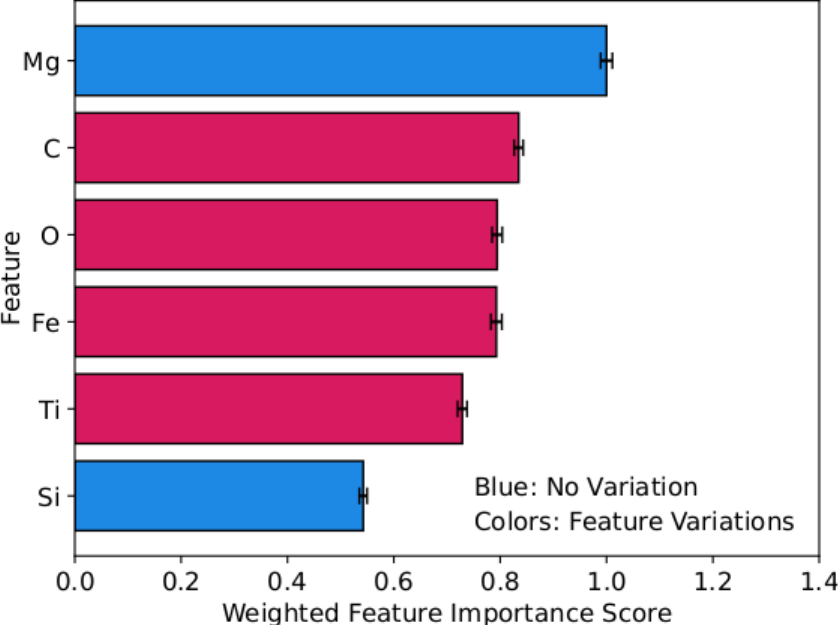}{0.5\textwidth}{(b) Ensemble 2}
          }
\gridline{\fig{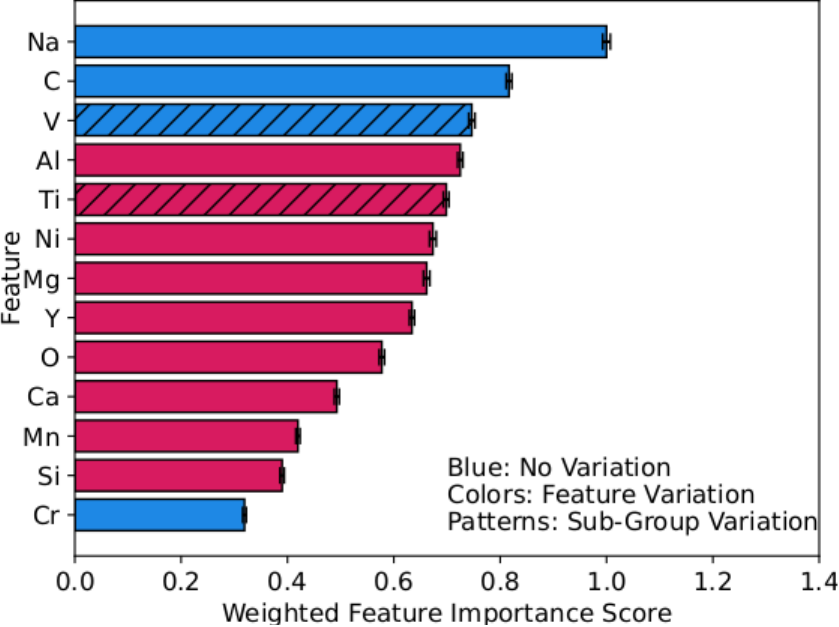}{0.5\textwidth}{(c) Ensemble 3}
          \fig{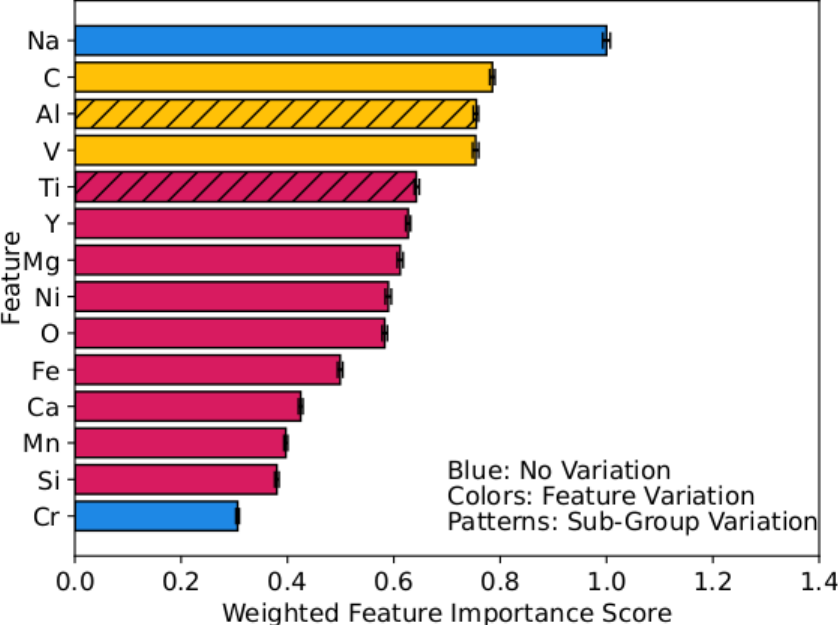}{0.5\textwidth}{(d) Ensemble 4}
          }
\caption{Same as Fig. \ref{fig:feature-importances-ex1} but for the abundance feature importance plots obtained for the Experiment 3 (super-Earth) ensembles. While C becomes the second most important feature in Ensemble 2 (b), Ensemble 3 (c) and Ensemble 4 (d), and the third most important feature in Ensemble 1 (a), this result is due to the use of null values within the ensembles as discussed in section \ref{subsec:nullvalues}. Therefore, the feature importance of C is not heavily considered. We note that Na remains the most important feature for Ensembles 3 (c) and 4 (d), while V remains near the top (although alternating) and Ni is now much further down. The elements V and Ti vary in importance for Ensemble 3 (c) while Al remains the fourth most important. A similar situation occurs where Al and Ti vary in importance for Ensemble 4 (d) while V remains the third most important. 
\label{fig:feature-importances-ex3-part1}}
\end{figure*}

\begin{figure*}[t]
\gridline{\fig{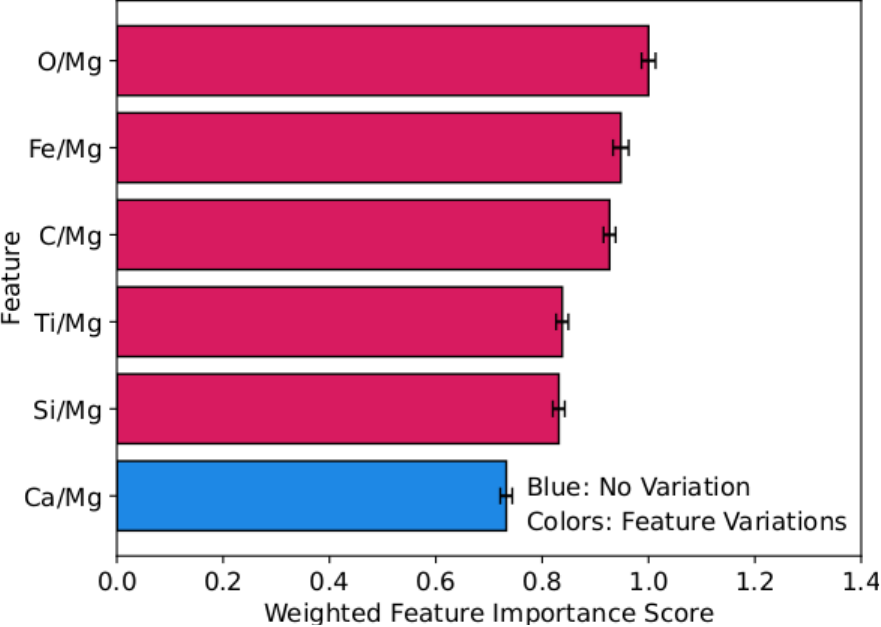}{0.5\textwidth}{(a) Ensemble 5}
          \fig{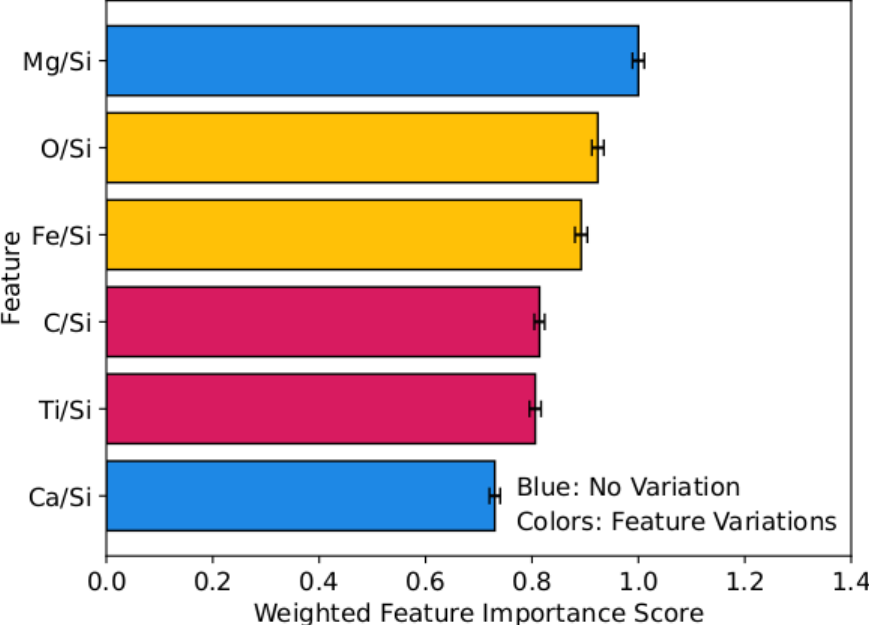}{0.5\textwidth}{(b) Ensemble 6}
          }
\gridline{\fig{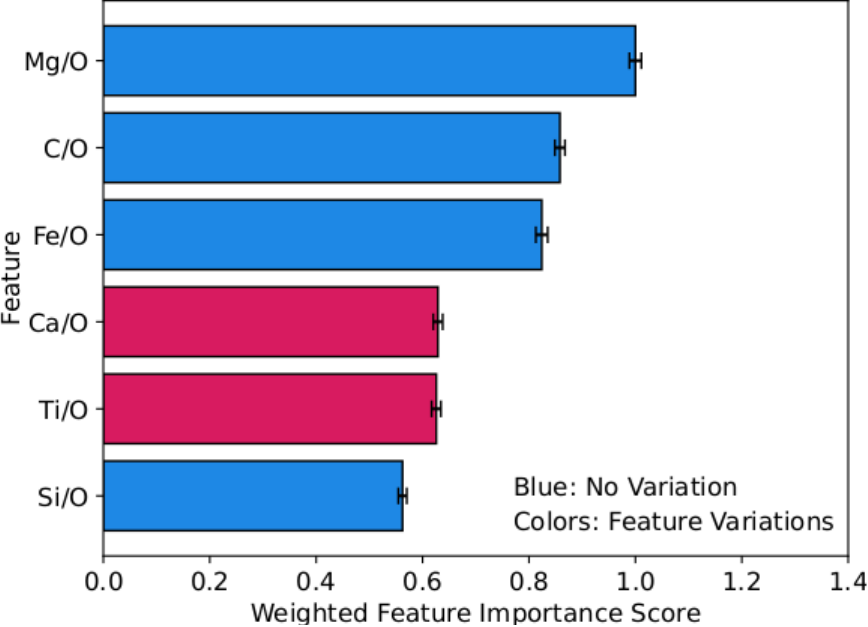}{0.5\textwidth}{(c) Ensemble 7}
          }
\caption{Similar to Fig. \ref{fig:feature-importances-ex1} but for the molar ratio feature importance plots obtained for the Experiment 3 (super-Earth) ensembles. As explained in Fig. \ref{fig:feature-importances-ex3-part1}, the feature importance for C was disregarded due to the use of null values. We note that most of the molar ratios included in Ensemble 5 (a) varied in terms of importance. 
\label{fig:feature-importances-ex3-part2}}
\end{figure*}

\subsubsection{Experiment 3: Super-Earths}\label{subsec:3}
The super-Earths include the planets that fall within the radii range of 1.0 $R_{\oplus}$ $<$ R$_{P}$ $<$ 2.0 $R_{\oplus}$, with 211 stars in the training sample.
Experiment 3 required including null values to run the algorithm successfully (see Section \ref{subsec:nullvalues}). This caused elements, such as C, to show high feature importance scores for the super-Earths. We note that C usually has the second or third most importance (Ensembles 1, 2, 3, and 4 -- see Fig. \ref{fig:feature-importances-ex3-part1}). The three elements with the highest null abundance values were C, O, and Y. Out of the 9,698 stars within the prediction sample, 3,408 stars did not include abundance values for C. For comparison, the elements O and Y had 3,603 and 5,167 null abundance values, respectively. Within our training sample of 211 stars, C only had 1 star without a recorded abundance, while O and Y had 2 and 4 stars, respectively, with null-valued abundances. Due to how XGBoost handles null valued abundances, the importance of C, O, and Y are not given much significance within our results. Once this is considered, we see that every ensemble show similar results to those seen for Experiment 1 and Experiment 2. As such, Mg remains the most important feature within Ensembles 1 and 2 (Fig. \ref{fig:feature-importances-ex3-part1}a and b, respectively) while Na retains the most importance for Ensembles 3 and 4. The element V alternates between being the third or fifth most important feature in Ensemble 3, while it is the fourth most important feature in Ensemble 4 (Fig. \ref{fig:feature-importances-ex3-part1}c and d, respectively). Ensemble 6 and 7 (see Fig. \ref{fig:feature-importances-ex3-part2}b and c, respectively) still retain Mg/Si and Mg/O, respectively, as the most important molar ratios.

\subsection{Abundance Comparisons}\label{subsec:abundance_plots}

\begin{figure*}[t]
\centering
\gridline{\fig{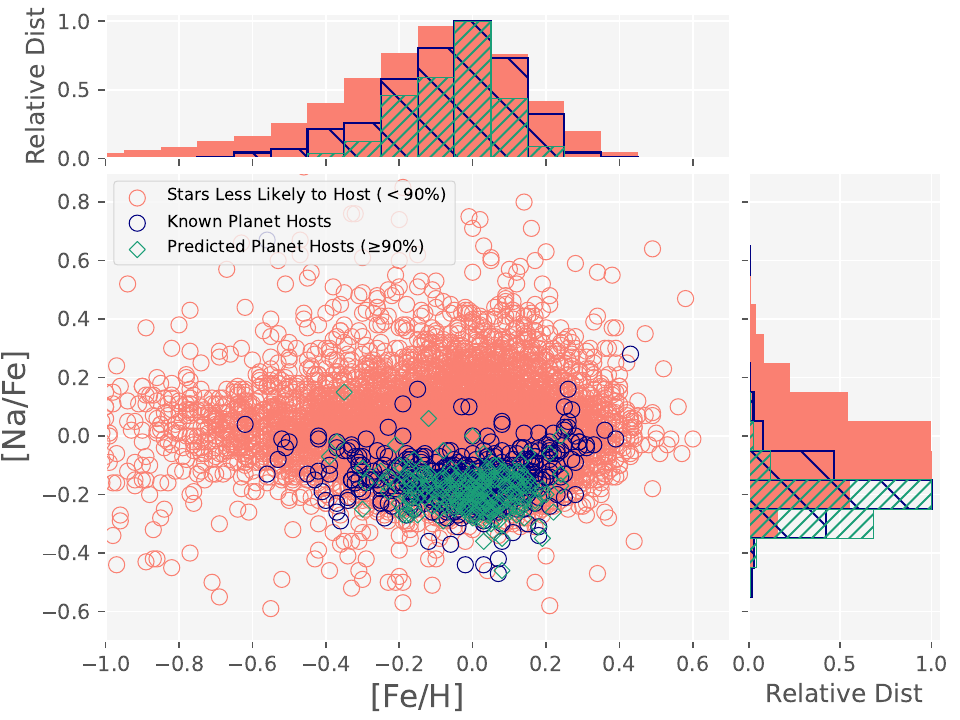}{0.54\textwidth}{(a) Small Planets}
          \fig{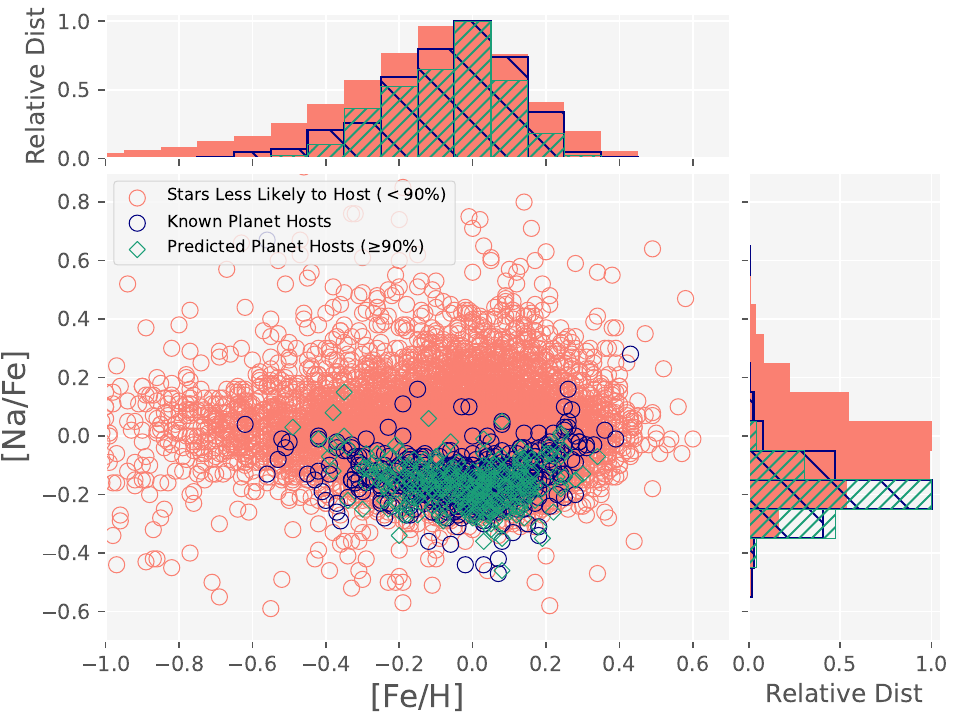}{0.54\textwidth}{(b) Sub-Neptunes}
          }
\gridline{\fig{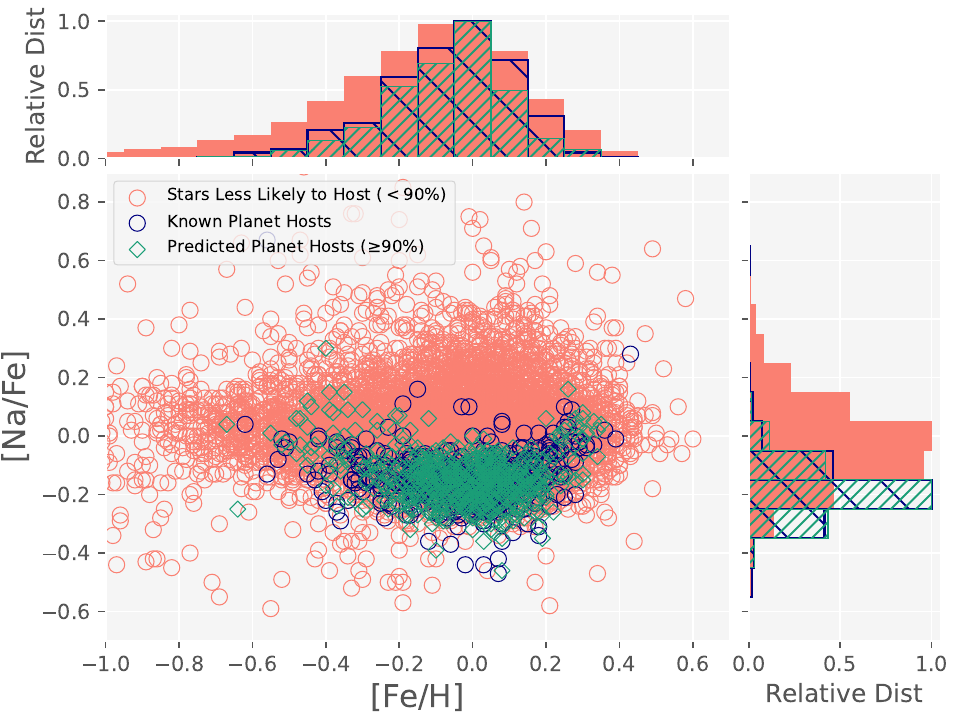}{0.54\textwidth}{(c) Super-Earths}
          }
\caption{The [Na/Fe] vs. [Fe/H] abundance plots (Ensemble 4) showing the stars that are most likely to host a planet (green), the known planet hosts (blue), and the stars less likely to host a planet (orange). Histograms are also included that show the [Na/Fe] relative distribution to the right of the scatterplot and [Fe/H] relative distribution at the top. We note the clear sub-solar Na abundances for each experiment's known and predicted planet hosts (see Section \ref{sec:discussion}). 
\label{fig:na-abundance-plots}}
\end{figure*}

We produced [Q/Fe] vs. [Fe/H] abundance plots to help understand how the stars that are unlikely to host exoplanets ($<$90$\%$; pink stars) compare to the known planet hosts (blue stars) and the predicted planet host abundances ($\geq$90$\%$, green stars) -- as shown in Fig. \ref{fig:na-abundance-plots}.
A threshold of $\geq$90$\%$ was chosen to indicate a model's prediction of a given star hosting a small planet (see Section \ref{subsec:golden}). The continual presence of Na at the top of the feature importance plots indicates that it might have a high significance in determining the presence of a small planet. We look at the [Na/Fe] vs. [Fe/H] abundance plots in Fig. \ref{fig:na-abundance-plots} to obtain additional insights into the significance of Na. The predicted hosts have sub-solar [Na/Fe] abundances in addition to solar-like [Fe/H] abundances for each experiment (Fig. \ref{fig:na-abundance-plots}a, b, and c, respectively). The one-dimensional histograms included at the right or top of each abundance plot allow us to visualize the distribution of stars in groupings of less likely, known, and predicted planet hosts based on their [Q/Fe] or [Fe/H] abundances. The colors of each group in the histogram match those from the scatter plot. From these histograms, we note that the [Na/Fe] abundances of the predicted hosts agree with the distribution of [Na/Fe] abundance observed for the known planet hosts (blue stars). This highly contrasts the [Na/Fe] abundance distributions for stars with $<$90$\%$ probability of hosting an exoplanet, which points to more solar-like abundances.

\begin{figure*}[t]
\centering
\gridline{\fig{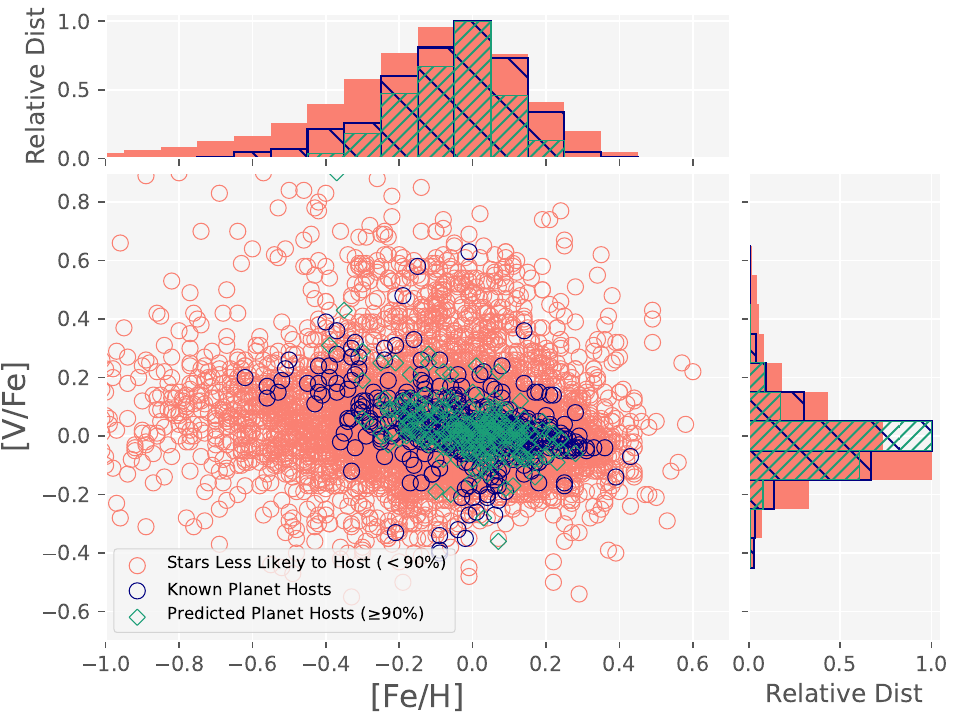}{0.54\textwidth}{(a) Small Planets}
\fig{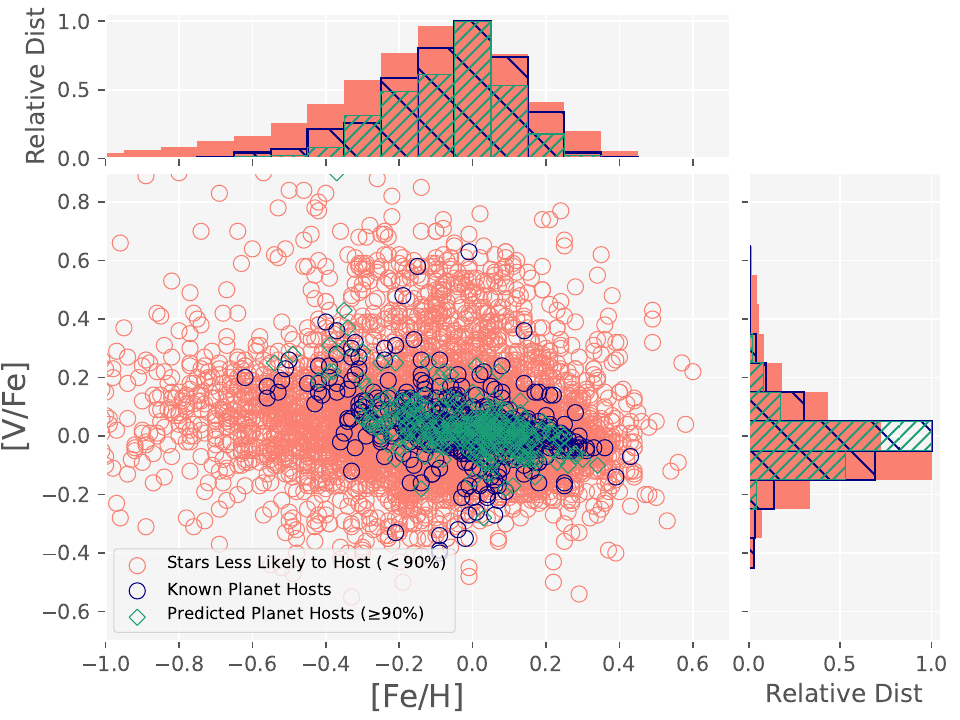}{0.54\textwidth}{(b) Sub-Neptunes} }
        
\gridline{\fig{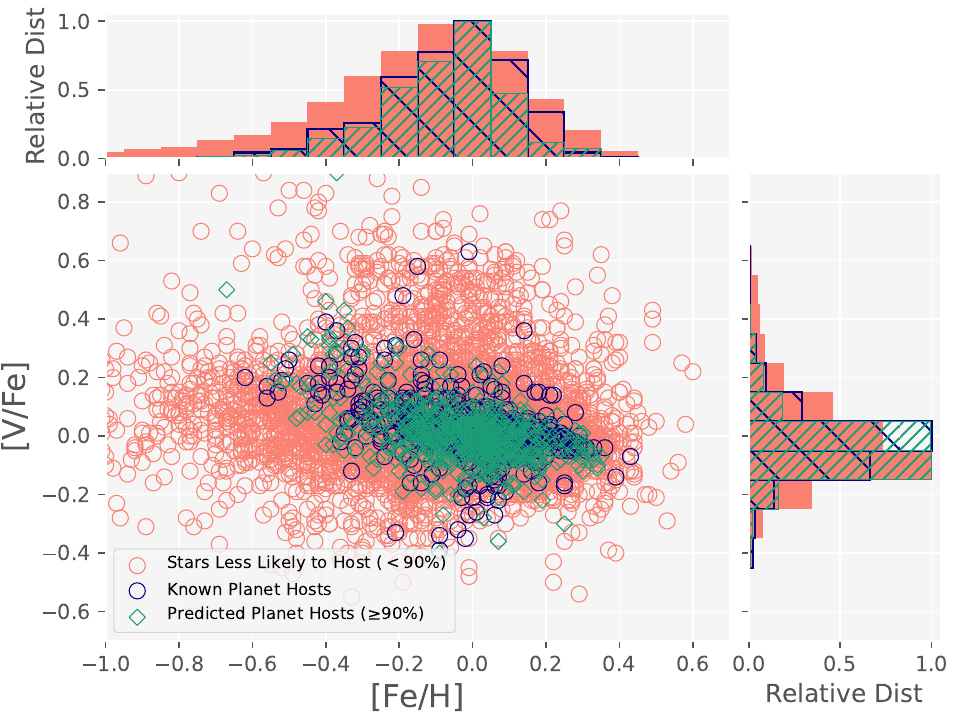}{0.54\textwidth}{(c) Super-Earths}
          }
\caption{Similar to Fig. \ref{fig:na-abundance-plots}, [V/Fe] vs. [Fe/H] abundance plots (Ensemble 4). 
\label{fig:v-abundance-plots}}
\end{figure*}

The element V was consistently shown to be one of the most important features for Ensemble 4 in each experiment. In the case of [V/Fe] (see Fig. \ref{fig:v-abundance-plots}), we see solar-like abundances for [V/Fe] and [Fe/H] for the predicted planet hosts. When examining the relative distribution histograms in Fig. \ref{fig:v-abundance-plots}, we note that the [Fe/H] abundances agree across all stellar groups: stars considered less likely to host, predicted hosts, and known hosts. However, the relative distributions of the [V/Fe] abundances for the predicted and known stellar hosts are more likely to be centered around solar-like abundances, while the stars less likely to host have [V/Fe] abundances with sub-solar values.

\subsection{Overlap of Predicted Host Stars Between Ensembles/Experiments} \label{subsec:overlaps}

\begin{figure*}[t]
\gridline{\fig{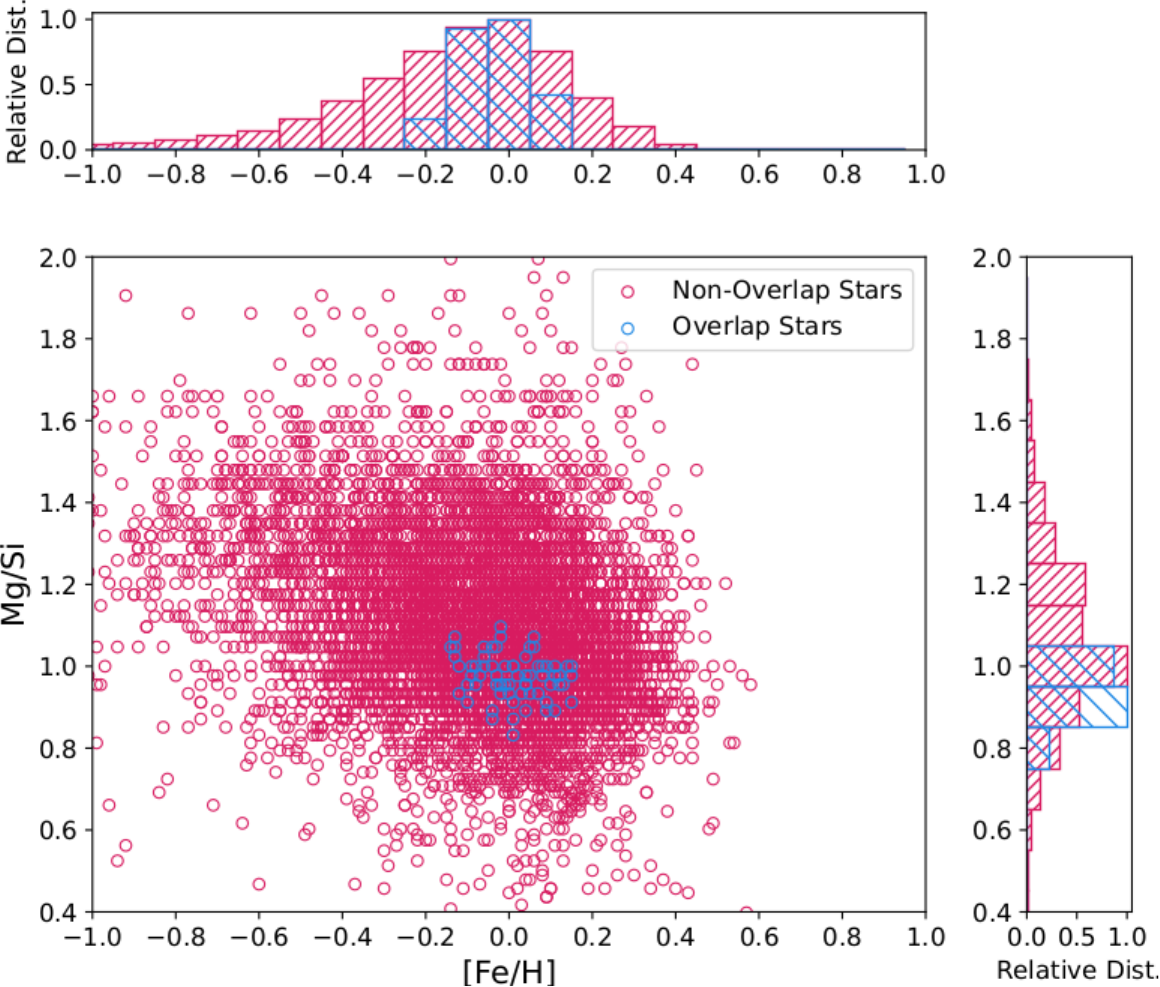}{0.54\textwidth}{(a) Small Planets}
          \fig{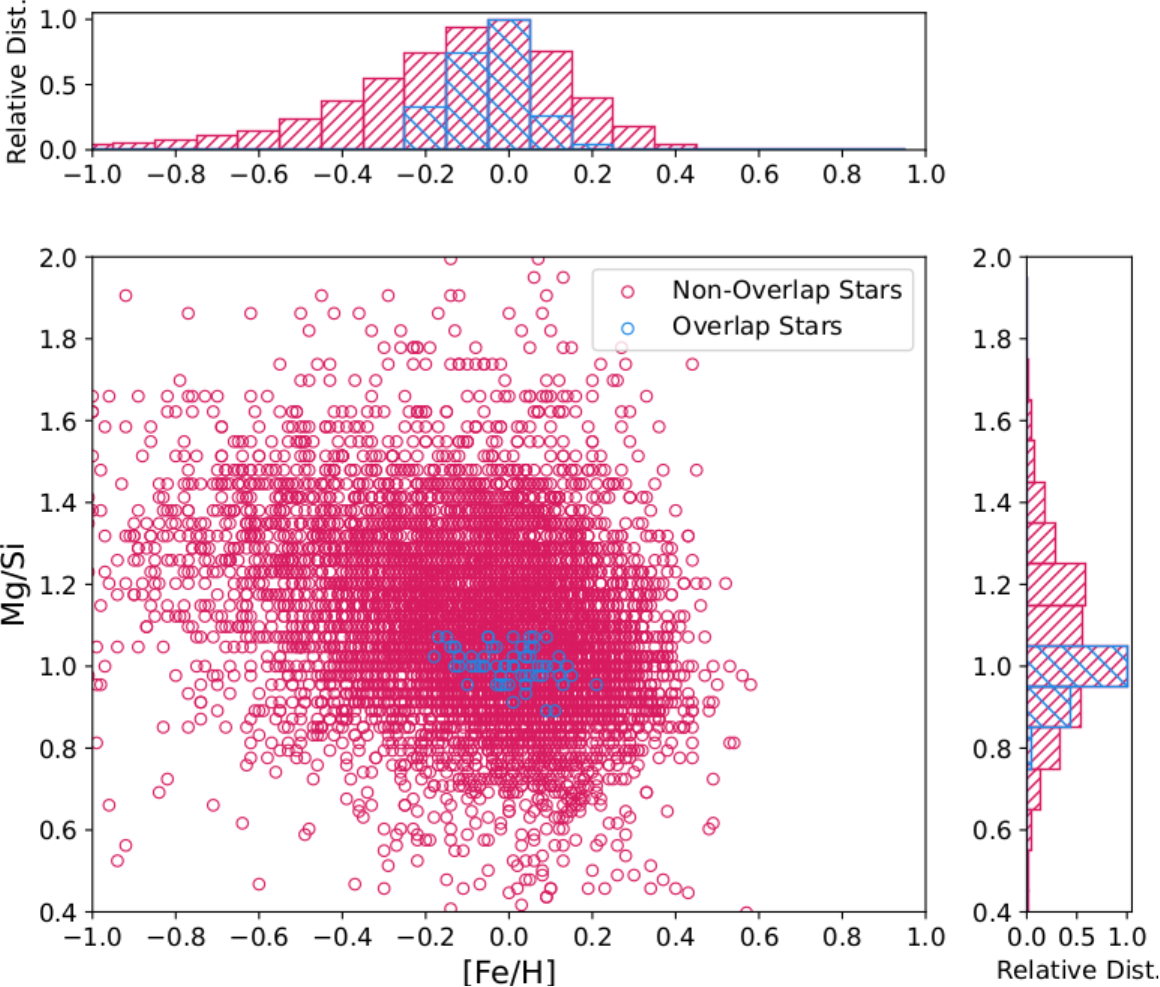}{0.54\textwidth}{(b) Sub-Neptunes}
          }
\gridline{\fig{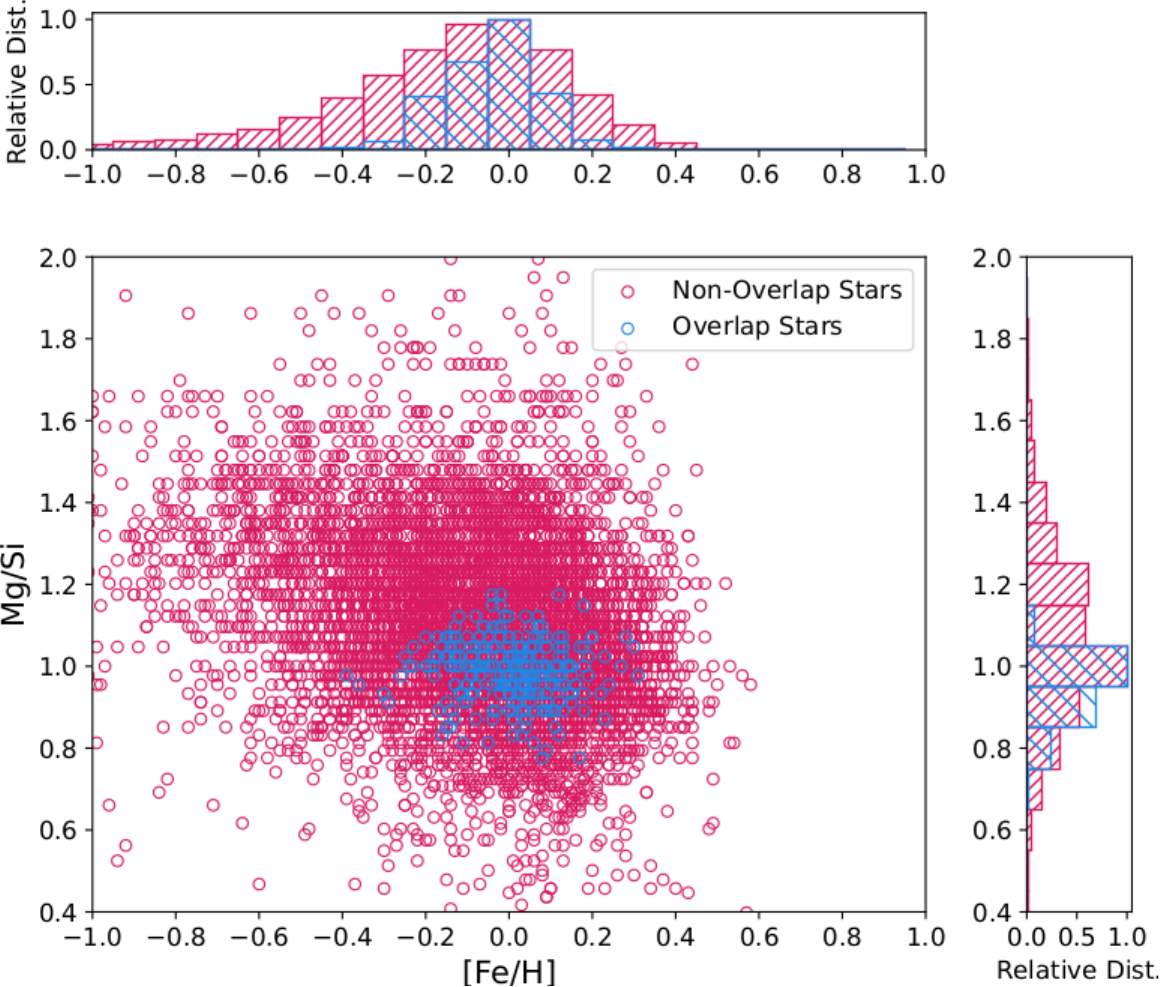}{0.54\textwidth}{(c) Super-Earths}
          }
\caption{The Mg/Si ratios of stars plotted against their corresponding [Fe/H] abundance. The overlap stars are shown in blue, while the prediction sample stars are plotted in red. The one-dimensional histograms of their [Fe/H] relative distributions are shown at the top of the scatterplot, the Mg/Si relative distributions are shown to the right. We note the relative distribution of Mg/Si of the small planets is concentrated at a ratio of $\approx$0.9. Once the subdivision of sub-Neptunes and super-Earth (b and c, respectively) was performed, the relative distribution of the Mg/Si of the overlap stars is $\approx$1.0. For all experiments, this implies that the Mg and Si throughout their systems are mostly present through olivines and pyroxenes \citep{Thiabaud2015a}.
\label{fig:mg/si-overlap}}
\end{figure*}

Overlap stars are defined as the predicted stellar hosts that have a $\geq$90$\%$ probability of hosting a small planet for each of the 7 ensembles in a given experiment. The overlap stars are defined as a crossmatch of all of the 21 lists (7 from each of the 3 experiments) with $\geq$90$\%$ prediction probability to determine which were the most predicted stellar hosts.
For each experiment, we obtained the following number of overlap stars: 67, 64, and 301, respectively (where the higher number of overlap stars for the super-Earths is due to the presence of null values). We plotted the overlap stars against those not consistently predicted as hosts for all experiments. To gain insight into the chemical ranges required for planet formation, we plotted the Mg/Si molar ratio of the overlap stars (shown in blue, see Fig. \ref{fig:mg/si-overlap}) against their [Fe/H] abundances, comparing them with the stars that are not consistently predicted as hosts (shown in red). We note that the Mg/Si ratios are primarily between 0.9 and 1.0 for 48$\%$, 30$\%$, and 34$\%$ of the overlap stars for Experiment 1, Experiment 2 and Experiment 3, respectively. The [Fe/H] abundances for the overlap stars are concentrated between -0.2 and 0.2 dex for Experiments 1 and 2, and between -0.4 and 0.4 dex for Experiment 3, respectively. The peak Mg/Si ratio distribution for Experiment 1 (Fig. \ref{fig:mg/si-overlap}a) is at a value of 0.9 while the Experiment 2 and Experiment 3 (Fig. \ref{fig:mg/si-overlap}b and c, respectively) Mg/Si ratio distribution peaks at 1.0. The implications of these Mg/Si ratios are further discussed in Section \ref{sec:discussion}. However, before considering insights from the overlap stars, we needed to test if they are statistically unique compared to those not consistently predicted as hosts for all experiments.

To test the statistical validity of the overlap stars being drawn from different distributions, we employed a two-dimensional, two-sample Kolmogorov-Smirnov test (2D KS-Test) using the \texttt{ndtest}\footnote{https://www.github.com/syrte/ndtest} package (which was built on the two-sided KS-Test approach highlighted in \citealt{Peacock1983,FasanoFranceschini1987,Press2007}). While it has been shown that the 2D KS-Test is accurate for a sample size of N $\gtrsim$ 20 and significance levels p $<$ 0.20 \citep{Press2007}, we adopt a more stringent definition for our sample size and significance levels, i.e. N $\geq$60, p $<$ 0.05 \citep{Gómez-de-Mariscal2021}. The two samples for this test were (a) the prediction sample and (b) the overlap stars, where the analysis was performed for each experiment. We compared the prediction sample stars [Fe/H] and [Q/H] abundances to the overlap stars [Fe/H] and [Q/H] abundances. This allows us to see how each element combination might affect each of the two samples. Results are shown in Table \ref{tab:ks-test}, where the ``N$_{pred}$" column contains the prediction sample stars, while the ``N$_{1}$", ``N$_{2}$", and ``N$_{3}$" columns represent the overlap stars for Experiments 1, 2, and 3 respectively. The ``p-value$_{1}$", ``p-value$_{2}$", and ``p-value$_{3}$" columns represent the p-value results for each experiment.

We also performed a 2D KS-Test comparing the disk locations for each star in our samples as determined by the Hypatia Catalog \citep{Bensby2003,Hinkel2014}. Three categories defined the disk location of a star in either the prediction sample or the overlap stars: N/A, thin, or thick disk, represented by the numbers 0, 1, or 2, respectively. The 2D KS-Test for the disk location was performed by comparing the prediction samples' [Fe/H] abundances and disk locations to the overlap stars' [Fe/H] abundances and disk locations. The p-value result for the 2D KS-Test of disk locations was $<<$ 0.05 in each experiment (see ``Disk" row in Table \ref{tab:ks-test}). This indicates that the disk locations for the overlap stars are drawn from different distributions, which makes them statistically unique. 

The p-values obtained for the disk location can be used in conjunction with the p-values of each element to gain insights into the required elements for small planet formation. If all elements have a p-value $<$ 0.05, then this would imply that the overlap stars show a unique make-up of elements that may indicate a small exoplanet's presence. As the overlap host stars have stellar abundances for multiple elements in a specific narrow range, that suggests that there might be an abundance-driven ``planet recipe" -- e.g., a narrow range of stellar abundances that are more likely to result in the presence of a small planet -- for detecting small planets, which may also link to planet formation. If some elements showed p-values $<$ 0.05 and are all thin disk stars, then the elements that had satisfactory p-values would likely be critical for planet formation. However, if some elements have p-values $<$ 0.05 and a mixture of thin/thick disk stars, then a closer look at the abundances between these stars would be performed. As shown in Table \ref{tab:ks-test}, each element obtained p-values $<$ 0.05. The 2D KS-Test, which analyzed disk locations for a star, also showed a p-value of $<$ 0.05. As such, we conclude that a unique planet recipe is suggested from these statistical tests. These results are further discussed in Section \ref{sec:discussion}.

\subsection{Additional Studies}\label{subsec:additional}
We intended to perform tests on other planetary parameters to understand how these could influence the feature importance scores. Testing on an eccentricity cutoff (e $\geq$ 0.2) would allow us to understand if there was a relationship between eccentricity and stellar [Fe/H] \citep{Dong-Sheng2023}. Separating our stellar sample between thin or thick disk stars would have allowed us to check if the features that determine the presence of small planets would differ between these systems \citep{Sahlholdt2021,Spitoni2023}. Finally we intended to create an experiment for planets between 0 $R_{\oplus}$ $<$ R$_{P}$ $<$ 1 $R_{\oplus}$ (low-radius planets) to obtain a complete picture on feature variations. However, in all of these cases, we did not have the necessary training sample for each test ($<$ 200 planets). These tests are left for a future study.

\section{Discussion} \label{sec:discussion}

The high feature importance scores obtained for Na in every experiment suggest that the element might play an important role during planetary formation. However, when comparing the compositions of the Sun and the Earth, there is a noticeable depletion of moderately volatile elements, such as Na, in the Earth's makeup \citep{Spaargaren2023}.
This depletion has been hypothesized to occur during the formation of rocky planetary bodies \citep{Wang2019}. It has also been shown to occur during steam atmosphere outgassing of planetesimals during their magma-ocean phase \citep{Fegley2016}. However, Na can still commonly be found in the crusts of Solar System bodies (Earth, Moon, Mars) in the form of plagioclases, e.g., $\textrm{(Ca,Na)}_{2}$(Al,Mg,Fe$^{2+}$)(Si$_{2}$O$_{7}$).

Additionally, as shown in Fig. \ref{fig:na-abundance-plots}, the [Na/Fe] abundance for both the known (blue) and predicted (green) planet hosts is mostly sub-solar. This indicates that while Na might be relevant for the formation of small planets, the quantities of Na necessary to start the formation process are low compared to the abundance levels found in the Sun. Additionally, the histograms show that the abundance trends of [Na/Fe] of known and predicted hosts are distinct from the trend shown for stars that are less likely to host planets (see Fig. \ref{fig:na-abundance-plots}). While this might be due to the limits in the detection methods for small planets, it could also hint at upper limits of Na abundances, which might hamper the formation of small planets. 

The significance of V in our results requires additional clarification. The element V can have various oxidation states (or the method by which electrons are lost, e.g., V$^{0}$, V$^{2+}$, V$^{3+}$, etc.) in planets consisting mainly of silicate rocks or metals \citep[e.g.][]{Sutton2005,Shearer2006,Qi2019}. The opposite process by which electrons are gained is called reduction. Geologists often obtain information related to oxidation and reduction (together referred to as redox) processes of mantle rocks on Earth (peridotites and basalts) through a dependence of V and the partial pressure of oxygen within mantle rocks, or the pressure oxygen would exert if it occupied the full volume of a mixture by itself at the same temperature \citep{Mallman2019}. As a result, it is possible to find a relationship between the quantity of V within mantle rock and the quantity of an element that has the same valence state as V \citep[e.g. Sc, Ga, Y or Zr,][]{Canil2002,Lee2005}. As such, it may be that the stellar abundance of V is an important feature because it is acting as a tracer for a different element that is more difficult to observe yet critical to the formation of small planets.

The element Ni showed significant importance for Ensemble 3 and Ensemble 4 in Experiment 1 and Experiment 2, respectively (see Fig. \ref{fig:feature-importances-ex1} and Fig. \ref{fig:feature-importances-ex2-part1}). \cite{Elardo2017} specifically note that chondritic abundances of Ni suggest that any planet or asteroid undergoing core formation is likely to incorporate Ni into its core, thereby undergoing isotopic fractionation. Therefore, Ni was a significant importance feature likely plays a role in the fractionation of Fe isotopes during planetary core formation.

The significance of Fe for small planets might be more evident through the analysis of the molar ratio ensembles. Ensemble 5 for Experiments 1 and 2 shows Fe/Mg being the most important molar ratio. In Experiment 3, while Fe/Mg (see Fig. \ref{fig:feature-importances-ex3-part2}a) appears to be a potentially significant feature, its importance exhibits considerable variation in relation to five of the other tested molar ratios. So while Fe/Mg may have the highest importance feature score at times, it is not consistent. As noted previously, Fe/Mg can influence the size of a planetary core and possibly determine the relative density of planets \citep{Hinkel2018, Schulze2021}. Therefore, Ensemble 5 for Experiments 1 and 2 could hint at the necessary Fe/Mg ratios in host stars required for the formation of planetary cores.

We found that Mg was the most important feature for Ensembles 1 and 2 in all three experiments. Given the nucleosynthetic origin of Fe and other $\alpha$-elements, a trend in Mg would normally be mirrored by similar trends in other $\alpha$-elements; therefore, the significance of Mg alone is notable, especially when considered as molar ratios. 
For example, for Experiment 3 in Ensemble 6, Mg/Si was the most important molar ratio. The Mg/Si ratios of the overlap stars in Experiment 3 range from 0.7 - 1.1 (see Fig. \ref{fig:mg/si-overlap}c). Only 14 stars shows Mg/Si $<$ 0.86, which indicates that Mg is more likely to form orthopyroxene while Si would be in the form of feldspars. The other 287 overlap stars show an Mg/Si ratio that ranges from 0.86 $<$ Mg/Si $<$ 1.73, which indicates that these systems likely have Mg and Si spread throughout olivines and pyroxenes \citep{Thiabaud2015a}. The Mg/Si ratios for the overlap stars suggest that super-Earth formation favors systems where silicates were distributed between olivine and pyroxenes in their protoplanetary disks. In the case of Experiment 1, while Mg/Si is shown as the most important feature in Ensemble 6 (see Fig. \ref{fig:molar-ratios-ex1}), this molar ratio varies significantly with Fe/Si and Ti/Si. Therefore, this could hint at the significance of Fe/Si or Ti/Si in determining the presence of a small planet for other classes of small planets.

In the case of Ensemble 7 for all three experiments, Mg/O was the molar ratio with the highest feature importance score. \cite{Harrison2018} indicated that this molar ratio could point to the distance from the progenitor star in which a white dwarf pollutant, e.g. exoplanet, exomoon, or asteroid that has been accreted onto the surface of a white dwarf, originally formed. They plotted the O/Mg abundance measurements for the white dwarf pollutants against formation distance, with an emphasis on temperature (see Figure 14). Three primary formation regions were identified: (a) depleted moderate volatile region -- where elements such as Mg, Si, Fe, Ni, and Cr are partially vaporized, (b) intermediate region -- which is depleted in volatiles such as Na, S, O, C, N, and (c) the volatile-rich region -- where ices can condense which increase O/Mg ratio. \cite{Harrison2018} noted that their models may occasionally be degenerate over a wide range of formation temperatures, but these models still generally indicate where the pollutants were formed.

Using the stellar abundances of systems with known exoplanets to predict potential small planet hosts revealed that the overlap stars' abundances for various elements were within a narrow range, hinting at a potential planet recipe for detecting small planets. Therefore, further constraining these abundance ranges might prove useful, as we could then target small planet detection missions at stars whose elemental makeup falls within these ranges to obtain a higher yield of small planet discoveries. Additionally, these abundances ranges might also imply the elemental abundances necessary to begin small planet formation. If so, this could allow us to determine which protoplanetary discs will likely form small planets. Thus, obtaining a planet recipe would allow us to clarify how small planets develop over time and aid the discovery of new exoplanets.

As more M-dwarf stellar abundances become available, we can constrain our study's results further as M-dwarf have advantages for detecting and characterizing small exoplanets \citep{Dressing15}. For example, we could be able to remove all host stars with null values and still have a sufficiently large sample within Experiment 3. Additionally, as lower radii planets are confirmed, this would allow us to train our data and predict potential hosts for planets between 0.0 $R_{\oplus}$ $<$ R$_P$ $<$ 1.0 $R_{\oplus}$ (see Section \ref{subsec:additional}).

We expect our results to highlight the importance of understanding the impact of elements such as Na and V in determining the presence of small planets based on the composition of a stellar host. Additionally, this study highlights the predictive capabilities of supervised classifiers in predicting likely exoplanet hosts. We postulate that the overlap stars obtained for this study are strong candidates for hosting small planets due to the results obtained from the p-value and XGBoost analyses. Therefore, the use of machine learning algorithms can be successful in creating target lists that are highly informative for current and future missions (such as JWST, NGRST, and HWO) dedicated to the detection and characterization of small planets.

\begin{acknowledgments}
The authors would like to acknowledge Dr. Angela Speck and Dr. Alan Whittington for their feedback and insight regarding the impacts of Na and V on planetary formation. NRH would like to thank Tatertot and Lasagna. L.G. acknowledges financial support from Fundação Carlos Chagas Filho de Amparo à Pesquisa do Estado do Rio de Janeiro (FAPERJ), through the ARC research grant E-26/211.386/2019 and an undergraduate fellowship. A.B. acknowledges financial support from Fundação Carlos Chagas Filho de Amparo à Pesquisa do Estado do Rio de Janeiro (FAPERJ) through an undergraduate fellowship. The research shown here acknowledges use of the Hypatia Catalog Database, an online compilation of stellar abundance data as described in \citep{Hinkel2014}. This research has made use of the NASA Exoplanet Archive, which is operated by the California Institute of Technology, under contract with the National Aeronautics and Space Administration under the Exoplanet Exploration Program. This research has made use of the VizieR catalogue access tool, CDS, Strasbourg, France (DOI : 10.26093/cds/vizier). The original description of the VizieR service was published in 2000, A$\&$AS 143, 23 \citep{Ochsenbein2000}. This research has made use of the SIMBAD database, operated at CDS, Strasbourg, France 2000, A$\&$AS, 143, 9 \citep{Wenger2000}.
\end{acknowledgments}

\newpage

\appendix


\begin{longtable*}{c c}
    \hline
    \hline
    Column Header & Description \\ [0.5ex]
    \hline
       Star Name & Stellar ID based on associated catalog \\
       Experiment & Experiment number for the star \\
       Overlap Star & Indicates if a star is an overlap star \\
       
       $\textrm{[Fe/H]}$ & [Fe/H] abundance in dex\\
       $\textrm{[C/H]}$ & [C/H] abundance in dex\\
       $\textrm{[O/H]}$ & [O/H] abundance in dex\\
       $\textrm{[Na/H]}$ & [Na/H] abundance in dex\\
       $\textrm{[Mg/H]}$ & [Mg/H] abundance in dex\\
       $\textrm{[Al/H]}$ & [Al/H] abundance in dex\\
       $\textrm{[Si/H]}$ & [Si/H] abundance in dex\\
       $\textrm{[Ca/H]}$ & [Ca/H] abundance in dex\\
       $\textrm{[Sc/H]}$ & [Sc/H] abundance in dex\\
       $\textrm{[Ti/H]}$ & [Ti/H] abundance in dex\\
       $\textrm{[V/H]}$ & [V/H] abundance in dex\\
       $\textrm{[Cr/H]}$ & [Cr/H] abundance in dex\\
       $\textrm{[Mn/H]}$ & [Mn/H] abundance in dex\\
       $\textrm{[Co/H]}$ & [Co/H] abundance in dex\\
       $\textrm{[Ni/H]}$ & [Ni/H] abundance in dex\\
       $\textrm{[Y/H]}$ & [Y/H] abundance in dex\\
       C/Mg & Molar ratio of C/Mg\\
       O/Mg & Molar ratio of O/Mg\\
       Si/Mg & Molar ratio of Si/Mg\\
       Ca/Mg & Molar ratio of Ca/Mg\\
       Ti/Mg & Molar ratio of Ti/Mg\\
       Fe/Mg & Molar ratio of Fe/Mg\\
       C/Si & Molar ratio of C/Si\\
       O/Si & Molar ratio of O/Si\\
       Mg/Si & Molar ratio of Mg/Si\\
       Ca/Si & Molar ratio of Ca/Si\\
       Ti/Si & Molar ratio of Ti/Si\\
       Fe/Si & Molar ratio of Fe/Si\\
       C/O & Molar ratio of C/O\\
       Si/O & Molar ratio of Si/O\\
       Mg/O & Molar ratio of Mg/O\\
       Ca/O & Molar ratio of Ca/O\\
       Ti/O & Molar ratio of Ti/O\\
       Fe/O & Molar ratio of Fe/O\\
       Exo & Indicates whether the star has an orbiting planet (0, 1, nan) \\
       Planet Letter & Planet ID, largest planet in the system\\
       Number of Planets & Number of Planets in system \\
       Planet Radius & Planet Radius (R$_{\bigoplus}$) \\
       Disk Location & Disk location of star (nan, 1, 2) \\     
       Ens1Prob & Ensemble 1 probability of hosting a small planet\\
       Ens2Prob & Ensemble 2 probability of hosting a small planet\\
       Ens3Prob & Ensemble 3 probability of hosting a small planet\\
       Ens4Prob & Ensemble 4 probability of hosting a small planet\\
       Ens5Prob & Ensemble 5 probability of hosting a small planet\\
       Ens6Prob & Ensemble 6 probability of hosting a small planet\\
       Ens7Prob & Ensemble 7 probability of hosting a small planet\\ 
   \caption{Description of the data available from this study. Stars that were trained and predicted upon for each experiment (1, 2, or 3) are indicated by the ``Experiment" value. For those overlap stars with $\geq90\%$ probability of hosting small planets across all ensembles and experiments (see Section \ref{subsec:overlaps}), the ``Overlap Star" column is set to ``Yes." For all stars, the abundances ([X/H] columns) and molar ratios (X/Y columns) are provided, in addition to likely disk location (``Disk Location"). For exoplanet hosting stars (``Exo" = 1), the ``Planet Letter", number of planets in a system (``Number of Planets"), and planet radius (``Planet Radius") are provided per the NASA Exoplanet Archive. In the last 7 columns, we include the probability scores of each star's likelihood of hosting a small planet. The table is 30,535 lines long, where stellar information is repeated between experiments. The full version of this table is available in an online machine-readable format.}\label{tab:main-list-1}
\end{longtable*}

\begin{center}
\begin{table*}[hbt!]
    \centering
    \begin{tabular}{c|c|c|c|c|c|c|c}
        \hline
        Feature & $N_{pred}$ & $N_{1}$ & $N_{2}$ & $N_{3}$  & p-value$_{1}$ & p-value$_{2}$ & p-value$_{3}$\\
        \hline
        Disk & 10178 & 67 & 64 & 301 & $2.14 \times 10^{-09}$ & $3.19 \times 10^{-07}$ & $2.38 \times 10^{-18}$\\
        
        Fe & 10178 & 67 & 64 & 301 & $2.77 \times 10^{-09}$ & $4.03 \times 10^{-07}$ & $3.93 \times 10^{-24}$\\
        
        C & 6768 & 67 & 64 & 301 & $5.48 \times 10^{-11}$ & $6.27 \times 10^{-12}$ & $2.20 \times 10^{-34}$\\
        
        O & 6570 & 67 & 64 & 301 & $5.34 \times 10^{-08}$ & $4.47 \times 10^{-08}$ & $2.14 \times 10^{-23}$\\
        
        Na & 7796 & 67 & 64 & 301 & $2.65 \times 10^{-15}$ & $1.00 \times 10^{-12}$ & $4.54 \times 10^{-41}$\\
        
        Mg & 8426 & 67 & 64 & 301 & $1.21 \times 10^{-13}$ & $1.22 \times 10^{-11}$ & $2.31 \times 10^{-38}$\\
        
        Al & 7549 & 67 & 64 & 301 & $5.84 \times 10^{-16}$ & $1.89 \times 10^{-13}$ & $8.00 \times 10^{-45}$\\
        
        Si & 7961 & 67 & 64 & 301 & $1.49 \times 10^{-11}$ & $2.64 \times 10^{-09}$ & $2.96 \times 10^{-31}$\\
        
        Ca & 8193 & 67 & 64 & 301 & $1.27 \times 10^{-09}$ & $9.82 \times 10^{-08}$ & $3.04 \times 10^{-24}$\\
        
        Ti & 8316 & 67 & 64 & 301 & $2.24 \times 10^{-11}$ & $5.24 \times 10^{-11}$ & $9.86 \times 10^{-33}$\\
        
        V & 7433 & 67 & 64 & 300 & $7.16 \times 10^{-12}$ & $9.69 \times 10^{-08}$ & $3.25 \times 10^{-30}$\\
        
        Cr & 7589 & 67 & 64 & 301 & $1.75 \times 10^{-10}$ & $6.48 \times 10^{-08}$ & $1.87 \times 10^{-25}$\\
        
        Mn & 7014 & 67 & 64 & 300 & $5.68 \times 10^{-11}$ & $2.62 \times 10^{-09}$ & $2.85 \times 10^{-29}$\\
        
        Ni & 7974 & 67 & 64 & 301 & $2.24 \times 10^{-10}$ & $9.91 \times 10^{-09}$ & $4.51 \times 10^{-26}$\\
        
        Y & 5004 & 67 & 64 & 301 & $2.37 \times 10^{-08}$ & $2.15 \times 10^{-06}$ & $1.22 \times 10^{-19}$\\

        \hline
    \end{tabular}
    \caption{2D KS-Test Results obtained for each experiment. The letter `N' represents the total number of stars available for each respective sample used in the 2D KS-Test analysis. The `pred' subscript represents the prediction sample of stars. Experiment 1 is represented by a subscript 1, Experiment 2 by a subscript 2, and Experiment 3 by a subscript 3 in their respective columns.}    \label{tab:ks-test}
\end{table*}
\end{center}


\vspace{5mm}

\begin{thebibliography}{}
\expandafter\ifx\csname natexlab\endcsname\relax\def\natexlab#1{#1}\fi
\providecommand{\url}[1]{\href{#1}{#1}}
\providecommand{\dodoi}[1]{doi:~\href{http://doi.org/#1}{\nolinkurl{#1}}}
\providecommand{\doeprint}[1]{\href{http://ascl.net/#1}{\nolinkurl{http://ascl.net/#1}}}
\providecommand{\doarXiv}[1]{\href{https://arxiv.org/abs/#1}{\nolinkurl{https://arxiv.org/abs/#1}}}

\bibitem[{{Alibert} {et~al.}(2013){Alibert}, {Carron}, {Fortier}, {Pfyffer}, {Benz}, {Mordasini}, \& {Swoboda}}]{Alibert2013}
{Alibert}, Y., {Carron}, F., {Fortier}, A., {et~al.} 2013, aap, 558, A109, \dodoi{10.1051/0004-6361/201321690}

\bibitem[{{An} {et~al.}(2023){An}, {Xie}, {Dai}, \& {Zhou}}]{Dong-Sheng2023}
{An}, D.-S., {Xie}, J.-W., {Dai}, Y.-Z., \& {Zhou}, J.-L. 2023, \aj, 165, 125, \dodoi{10.3847/1538-3881/acb533}

\bibitem[{{Bashi} \& {Zucker}(2019)}]{Bashi2019}
{Bashi}, D., \& {Zucker}, S. 2019, \aj, 158, 61, \dodoi{10.3847/1538-3881/ab27c9}

\bibitem[{{Bensby} {et~al.}(2003){Bensby}, {Feltzing}, \& {Lundstr{\"o}m}}]{Bensby2003}
{Bensby}, T., {Feltzing}, S., \& {Lundstr{\"o}m}, I. 2003, \aap, 410, 527, \dodoi{10.1051/0004-6361:20031213}

\bibitem[{{Bergsten} {et~al.}(2022){Bergsten}, {Pascucci}, {Mulders}, {Fernandes}, \& {Koskinen}}]{Bergsten2022}
{Bergsten}, G.~J., {Pascucci}, I., {Mulders}, G.~D., {Fernandes}, R.~B., \& {Koskinen}, T.~T. 2022, \aj, 164, 190, \dodoi{10.3847/1538-3881/ac8fea}

\bibitem[{{Bond} {et~al.}(2010){Bond}, {O'Brien}, \& {Lauretta}}]{Bond2010}
{Bond}, J.~C., {O'Brien}, D.~P., \& {Lauretta}, D.~S. 2010, \apj, 715, 1050, \dodoi{10.1088/0004-637X/715/2/1050}

\bibitem[{{Bonsor} {et~al.}(2011){Bonsor}, {Mustill}, \& {Wyatt}}]{Bonsor2011}
{Bonsor}, A., {Mustill}, A.~J., \& {Wyatt}, M.~C. 2011, \mnras, 414, 930, \dodoi{10.1111/j.1365-2966.2011.18524.x}

\bibitem[{{Brewer} {et~al.}(2018){Brewer}, {Wang}, {Fischer}, \& {Foreman-Mackey}}]{Brewer2018}
{Brewer}, J.~M., {Wang}, S., {Fischer}, D.~A., \& {Foreman-Mackey}, D. 2018, \apjl, 867, L3, \dodoi{10.3847/2041-8213/aae710}

\bibitem[{Canil(2002)}]{Canil2002}
Canil, D. 2002, Earth and Planetary Science Letters, 195, 75

\bibitem[{Chen \& Guestrin(2016)}]{ChenandGuestrin2016}
Chen, T., \& Guestrin, C. 2016, CoRR, abs/1603.02754

\bibitem[{{Dressing} \& {Charbonneau}(2015)}]{Dressing15}
{Dressing}, C.~D., \& {Charbonneau}, D. 2015, \apj, 807, 45, \dodoi{10.1088/0004-637X/807/1/45}

\bibitem[{{Elardo} \& {Shahar}(2017)}]{Elardo2017}
{Elardo}, S.~M., \& {Shahar}, A. 2017, Nature Geoscience, 10, 317, \dodoi{10.1038/ngeo2896}

\bibitem[{{Fasano} \& {Franceschini}(1987)}]{FasanoFranceschini1987}
{Fasano}, G., \& {Franceschini}, A. 1987, mnras, 225, 155, \dodoi{10.1093/mnras/225.1.155}

\bibitem[{Fegley {et~al.}(2016)Fegley, Jacobson, Williams, Plane, Schaefer, \& Lodders}]{Fegley2016}
Fegley, B., Jacobson, N.~S., Williams, K., {et~al.} 2016, The Astrophysical Journal, 824, 103

\bibitem[{Feng {et~al.}(2021)Feng, Zhou, \& Tong}]{Feng21}
Feng, Y., Zhou, M., \& Tong, X. 2021, Statistical Analysis and Data Mining: The ASA Data Science Journal, 14, 383, \dodoi{https://doi.org/10.1002/sam.11538}

\bibitem[{Fischer \& Valenti(2005)}]{Fischer05}
Fischer, D., \& Valenti, J. 2005, ApJ, 622, 1102

\bibitem[{{Fulton} {et~al.}(2017){Fulton}, {Petigura}, {Howard}, {Isaacson}, {Marcy}, {Cargile}, {Hebb}, {Weiss}, {Johnson}, {Morton}, {Sinukoff}, {Crossfield}, \& {Hirsch}}]{Fulton2017}
{Fulton}, B.~J., {Petigura}, E.~A., {Howard}, A.~W., {et~al.} 2017, \aj, 154, 109, \dodoi{10.3847/1538-3881/aa80eb}

\bibitem[{{Gaillard} {et~al.}(2022){Gaillard}, {Bernadou}, {Roskosz}, {Bouhifd}, {Marrocchi}, {Iacono-Marziano}, {Moreira}, {Scaillet}, \& {Rogerie}}]{Gaillard2022}
{Gaillard}, F., {Bernadou}, F., {Roskosz}, M., {et~al.} 2022, Earth and Planetary Science Letters, 577, 117255, \dodoi{10.1016/j.epsl.2021.117255}

\bibitem[{G{\'o}mez-de Mariscal {et~al.}(2021)G{\'o}mez-de Mariscal, Guerrero, Sneider, Jayatilaka, Phillip, Wirtz, \& Mu{\~{n}}oz-Barrutia}]{Gómez-de-Mariscal2021}
G{\'o}mez-de Mariscal, E., Guerrero, V., Sneider, A., {et~al.} 2021, Scientific Reports, 11, 20942, \dodoi{10.1038/s41598-021-00199-5}

\bibitem[{{Gonzalez}(1997)}]{Gonzalez1997}
{Gonzalez}, G. 1997, \mnras, 285, 403, \dodoi{10.1093/mnras/285.2.403}

\bibitem[{{Gonzalez}(2009)}]{Gonzalez2009}
---. 2009, \mnras, 399, L103, \dodoi{10.1111/j.1745-3933.2009.00734.x}

\bibitem[{{Harrison} {et~al.}(2018){Harrison}, {Bonsor}, \& {Madhusudhan}}]{Harrison2018}
{Harrison}, J. H.~D., {Bonsor}, A., \& {Madhusudhan}, N. 2018, mnras, 479, 3814, \dodoi{10.1093/mnras/sty1700}

\bibitem[{{Helled} {et~al.}(2014){Helled}, {Bodenheimer}, {Podolak}, {Boley}, {Meru}, {Nayakshin}, {Fortney}, {Mayer}, {Alibert}, \& {Boss}}]{Helled2014}
{Helled}, R., {Bodenheimer}, P., {Podolak}, M., {et~al.} 2014, in Protostars and Planets VI, ed. H.~{Beuther}, R.~S. {Klessen}, C.~P. {Dullemond}, \& T.~{Henning}, 643--665, \dodoi{10.2458/azu_uapress_9780816531240-ch028}

\bibitem[{{Hinkel} {et~al.}(2014){Hinkel}, {Timmes}, {Young}, {Pagano}, \& {Turnbull}}]{Hinkel2014}
{Hinkel}, N.~R., {Timmes}, F.~X., {Young}, P.~A., {Pagano}, M.~D., \& {Turnbull}, M.~C. 2014, \aj, 148, 54, \dodoi{10.1088/0004-6256/148/3/54}

\bibitem[{{Hinkel} {et~al.}(2019){Hinkel}, {Unterborn}, {Kane}, {Somers}, \& {Galvez}}]{Hinkel2019}
{Hinkel}, N.~R., {Unterborn}, C., {Kane}, S.~R., {Somers}, G., \& {Galvez}, R. 2019, \apj, 880, 49, \dodoi{10.3847/1538-4357/ab27c0}

\bibitem[{{Hinkel} \& {Unterborn}(2018)}]{Hinkel2018}
{Hinkel}, N.~R., \& {Unterborn}, C.~T. 2018, \apj, 853, 83, \dodoi{10.3847/1538-4357/aaa5b4}

\bibitem[{{Hinkel} {et~al.}(2022){Hinkel}, {Young}, \& {Wheeler}}]{Hinkel2022}
{Hinkel}, N.~R., {Young}, P.~A., \& {Wheeler}, Caleb~H., I. 2022, \aj, 164, 256, \dodoi{10.3847/1538-3881/ac9bfa}

\bibitem[{Lee {et~al.}(2005)Lee, Leeman, Canil, \& Li}]{Lee2005}
Lee, C.-T., Leeman, W., Canil, D., \& Li, Z. 2005, Geochimica et Cosmochimica Acta Supplement, 69, A639

\bibitem[{{Liu} \& {Ji}(2020)}]{Liu2020}
{Liu}, B., \& {Ji}, J. 2020, Research in Astronomy and Astrophysics, 20, 164, \dodoi{10.1088/1674-4527/20/10/164}

\bibitem[{{Loaiza-Tacuri} {et~al.}(2023){Loaiza-Tacuri}, {Cunha}, {Smith}, {Martinez}, {Ghezzi}, {Schuler}, {Teske}, \& {Howell}}]{Loaiza-Tacuri2023}
{Loaiza-Tacuri}, V., {Cunha}, K., {Smith}, V.~V., {et~al.} 2023, \apj, 946, 61, \dodoi{10.3847/1538-4357/acb137}

\bibitem[{Mallmann \& O’Neill(2009)}]{Mallman2019}
Mallmann, G., \& O’Neill, H. S.~C. 2009, Journal of Petrology, 50, 1765, \dodoi{10.1093/petrology/egp053}

\bibitem[{{Marboeuf} {et~al.}(2014{\natexlab{a}}){Marboeuf}, {Thiabaud}, {Alibert}, {Cabral}, \& {Benz}}]{Marboeuf2014a}
{Marboeuf}, U., {Thiabaud}, A., {Alibert}, Y., {Cabral}, N., \& {Benz}, W. 2014{\natexlab{a}}, \aap, 570, A35, \dodoi{10.1051/0004-6361/201322207}

\bibitem[{{Marboeuf} {et~al.}(2014{\natexlab{b}}){Marboeuf}, {Thiabaud}, {Alibert}, {Cabral}, \& {Benz}}]{Marboeuf2014b}
---. 2014{\natexlab{b}}, \aap, 570, A36, \dodoi{10.1051/0004-6361/201423431}

\bibitem[{{Martinez} {et~al.}(2019){Martinez}, {Cunha}, {Ghezzi}, \& {Smith}}]{Martinez2019}
{Martinez}, C.~F., {Cunha}, K., {Ghezzi}, L., \& {Smith}, V.~V. 2019, \apj, 875, 29, \dodoi{10.3847/1538-4357/ab0d93}

\bibitem[{McDonough(2003)}]{McDonough2003}
McDonough, W.~F. 2003, 2.15 - Compositional Model for the Earth's Core (Oxford: Pergamon), 547--568, \dodoi{https://doi.org/10.1016/B0-08-043751-6/02015-6}

\bibitem[{{Ochsenbein} {et~al.}(2000){Ochsenbein}, {Bauer}, \& {Marcout}}]{Ochsenbein2000}
{Ochsenbein}, F., {Bauer}, P., \& {Marcout}, J. 2000, aaps, 143, 23, \dodoi{10.1051/aas:2000169}

\bibitem[{{Peacock}(1983)}]{Peacock1983}
{Peacock}, J.~A. 1983, mnras, 202, 615, \dodoi{10.1093/mnras/202.3.615}

\bibitem[{Press(2007)}]{Press2007}
Press, W. 2007, Numerical Recipes 3rd Edition: The Art of Scientific Computing, Numerical Recipes: The Art of Scientific Computing (Cambridge University Press).
\newblock \url{https://books.google.com/books?id=1aAOdzK3FegC}

\bibitem[{{Qi} {et~al.}(2019){Qi}, {Wu}, {Ionov}, {Puchtel}, {Carlson}, {Nicklas}, {Yu}, {Kang}, {Li}, \& {Huang}}]{Qi2019}
{Qi}, Y.-H., {Wu}, F., {Ionov}, D.~A., {et~al.} 2019, \gca, 259, 288, \dodoi{10.1016/j.gca.2019.06.008}

\bibitem[{Sahlholdt {et~al.}(2021)Sahlholdt, Feltzing, \& Feuillet}]{Sahlholdt2021}
Sahlholdt, C.~L., Feltzing, S., \& Feuillet, D.~K. 2021, Monthly Notices of the Royal Astronomical Society, 510, 4669, \dodoi{10.1093/mnras/stab3681}

\bibitem[{{Schulze} {et~al.}(2021){Schulze}, {Wang}, {Johnson}, {Gaudi}, {Unterborn}, \& {Panero}}]{Schulze2021}
{Schulze}, J.~G., {Wang}, J., {Johnson}, J.~A., {et~al.} 2021, \psj, 2, 113, \dodoi{10.3847/PSJ/abcaa8}

\bibitem[{Shearer {et~al.}(2006)Shearer, McKay, Papike, \& Karner}]{Shearer2006}
Shearer, C., McKay, G., Papike, J., \& Karner, J. 2006, American Mineralogist, 91, 1657

\bibitem[{{Smyth} {et~al.}(2006){Smyth}, {Frost}, {Nestola}, {Holl}, \& {Bromiley}}]{Smyth2006}
{Smyth}, J.~R., {Frost}, D.~J., {Nestola}, F., {Holl}, C.~M., \& {Bromiley}, G. 2006, \grl, 33, L15301, \dodoi{10.1029/2006GL026194}

\bibitem[{{Spaargaren} {et~al.}(2023){Spaargaren}, {Wang}, {Mojzsis}, {Ballmer}, \& {Tackley}}]{Spaargaren2023}
{Spaargaren}, R.~J., {Wang}, H.~S., {Mojzsis}, S.~J., {Ballmer}, M.~D., \& {Tackley}, P.~J. 2023, \apj, 948, 53, \dodoi{10.3847/1538-4357/acac7d}

\bibitem[{{Spitoni, E.} {et~al.}(2023){Spitoni, E.}, {Recio-Blanco, A.}, {de Laverny, P.}, {Palicio, P. A.}, {Kordopatis, G.}, {Schultheis, M.}, {Contursi, G.}, {Poggio, E.}, {Romano, D.}, \& {Matteucci, F.}}]{Spitoni2023}
{Spitoni, E.}, {Recio-Blanco, A.}, {de Laverny, P.}, {et~al.} 2023, A\&A, 670, A109, \dodoi{10.1051/0004-6361/202244349}

\bibitem[{Sutton {et~al.}(2005)Sutton, Karner, Papike, Delaney, Shearer, Newville, Eng, Rivers, \& Dyar}]{Sutton2005}
Sutton, S., Karner, J., Papike, J., {et~al.} 2005, Geochimica et Cosmochimica Acta, 69, 2333

\bibitem[{{Tamayo} {et~al.}(2016){Tamayo}, {Silburt}, {Valencia}, {Menou}, {Ali-Dib}, {Petrovich}, {Huang}, {Rein}, {van Laerhoven}, {Paradise}, {Obertas}, \& {Murray}}]{Tamayo2016}
{Tamayo}, D., {Silburt}, A., {Valencia}, D., {et~al.} 2016, apjl, 832, L22, \dodoi{10.3847/2041-8205/832/2/L22}

\bibitem[{{Thiabaud} {et~al.}(2014){Thiabaud}, {Marboeuf}, {Alibert}, {Cabral}, {Leya}, \& {Mezger}}]{Thiabaud2014}
{Thiabaud}, A., {Marboeuf}, U., {Alibert}, Y., {et~al.} 2014, aap, 562, A27, \dodoi{10.1051/0004-6361/201322208}

\bibitem[{{Thiabaud} {et~al.}(2015{\natexlab{a}}){Thiabaud}, {Marboeuf}, {Alibert}, {Leya}, \& {Mezger}}]{Thiabaud2015a}
{Thiabaud}, A., {Marboeuf}, U., {Alibert}, Y., {Leya}, I., \& {Mezger}, K. 2015{\natexlab{a}}, \aap, 580, A30, \dodoi{10.1051/0004-6361/201525963}

\bibitem[{{Thiabaud} {et~al.}(2015{\natexlab{b}}){Thiabaud}, {Marboeuf}, {Alibert}, {Leya}, \& {Mezger}}]{Thiabaud2015b}
---. 2015{\natexlab{b}}, aap, 574, A138, \dodoi{10.1051/0004-6361/201424868}

\bibitem[{{Unterborn} {et~al.}(2023){Unterborn}, {Desch}, {Haldemann}, {Lorenzo}, {Schulze}, {Hinkel}, \& {Panero}}]{Unterborn2023}
{Unterborn}, C.~T., {Desch}, S.~J., {Haldemann}, J., {et~al.} 2023, \apj, 944, 42, \dodoi{10.3847/1538-4357/acaa3b}

\bibitem[{{Wang} {et~al.}(2019){Wang}, {Lineweaver}, \& {Ireland}}]{Wang2019}
{Wang}, H.~S., {Lineweaver}, C.~H., \& {Ireland}, T.~R. 2019, \icarus, 328, 287, \dodoi{10.1016/j.icarus.2019.03.018}

\bibitem[{{Wenger} {et~al.}(2000){Wenger}, {Ochsenbein}, {Egret}, {Dubois}, {Bonnarel}, {Borde}, {Genova}, {Jasniewicz}, {Lalo{\"e}}, {Lesteven}, \& {Monier}}]{Wenger2000}
{Wenger}, M., {Ochsenbein}, F., {Egret}, D., {et~al.} 2000, \aaps, 143, 9, \dodoi{10.1051/aas:2000332}

\bibitem[{{Xu} {et~al.}(2019){Xu}, {Dufour}, {Klein}, {Melis}, {Monson}, {Zuckerman}, {Young}, \& {Jura}}]{Xu2019}
{Xu}, S., {Dufour}, P., {Klein}, B., {et~al.} 2019, \aj, 158, 242, \dodoi{10.3847/1538-3881/ab4cee}

\end{thebibliography}
\end{document}